\documentclass[nofootinbib,twocolumn,a4paper]{revtex4}
\usepackage[]{graphicx}
\usepackage{amsfonts}
\usepackage{amssymb}
\usepackage{amsmath}
\usepackage{latexsym}
\usepackage{subfloat}

\begin{document}

\preprint{ICG preprint 04/xx}

\title{Braneworld resonances}

\author{Chris Clarkson}
\email{chris.clarkson@port.ac.uk}%
\author{Sanjeev S.~Seahra}
\email{sanjeev.seahra@port.ac.uk}%
\affiliation{Institute of Cosmology \& Gravitation, University of
Portsmouth, Portsmouth, PO1 2EG, UK}

\setlength\arraycolsep{2pt}
\newcommand*{\di}{\partial}
\newcommand*{\V}{{\mathcal V}^{(k)}_d}
\newcommand*{\volume}{\sqrt{\sigma^{(k,d)}}}
\newcommand*{\OneTwo}{{(1,2)}}
\newcommand*{\onetwo}{{1,2}}
\newcommand*{\Lm}{{\mathcal L}_m}
\newcommand*{\stm}{{\textsc{stm}}}
\newcommand*{\Hm}{{\mathcal H}_m}
\newcommand*{\hatHm}{\hat{\mathcal H}_m}
\newcommand*{\Ldust}{{\mathcal L}_\mathrm{dust}}
\newcommand*{\maxsym}{{\mathbb S}_d^{(k)}}
\newcommand*{\sn}{{\mathrm{sn}}}
\newcommand*{\cn}{{\mathrm{cn}}}
\newcommand*{\nc}{{\mathrm{nc}}}
\newcommand*{\Jacobisc}{{\mathrm{sc}}}
\newcommand*{\ansatz}{{\emph{ansatz}}}
\newcommand*{\ds}[1]{ds^2_\text{\tiny{($#1$)}}}
\newcommand*{\kret}[1]{\mathfrak{K}_\text{\tiny{($#1$)}}}
\newcommand*{\ads}[1]{{AdS$_{#1}$}}
\newcommand*{\rhotot}{\rho_\text{tot}}
\newcommand*{\ptot}{p_\text{tot}}
\newcommand*{\rb}{{r_\mathrm{b}}}
\newcommand*{\fb}{{f_\mathrm{b}}}
\newcommand*{\xb}{{x_\text{b}}}
\newcommand*{\Zb}{{Z_\text{b}}}
\newcommand*{\xp}{{x_\text{p}}}
\newcommand*{\zb}{{z_\text{b}}}
\newcommand*{\Rh}{{R_\text{h}}}
\newcommand*{\psib}{{\psi_\text{b}}}
\newcommand*{\Imag}{\text{Im}\,}
\newcommand*{\Real}{\text{Re}\,}
\newcommand*{\x}{X}
\newcommand*{\qnm}{\text{\textsc{QNM}}}
\newcommand*{\omegaGR}{\omega_\text{\textsc{GR}}}
\newcommand*{\PsiL}{\Psi^\text{\tiny (L)}_\omega}
\newcommand*{\PsiR}{\Psi^\text{\tiny (R)}_\omega}
\newcommand*{\Psiout}{\Psi^\text{\tiny (out)}_\omega}

\date{\today}

\begin{abstract}

We investigate in detail gravitational waves in an
Schwarzschild-anti-de Sitter bulk spacetime surrounded by an
Einstein static brane with generic matter content.  Such a model
provides a useful analogy to braneworld cosmology at various
stages of its evolution, and generalizes our previous work
[gr-qc/0504023] on pure tension Einstein-static branes.  We find
that the behaviour of tensor-mode perturbations is completely
dominated by quasi-normal modes, and we use a variety of numeric
and analytic techniques to find the frequencies and lifetimes of
these excitations.  The parameter space governing the model yields
a rich variety of resonant phenomena, which we thoroughly explore.
We find that certain configurations can support a number of
lightly damped `quasi-bound states'. A zero-mode which reproduces
4-dimensional general relativity is recovered on infinitely large
branes.  We also examine the problem in the time domain using
Green's function techniques in addition to direct numeric
integration. We conclude by discussing how the quasi-normal
resonances we find here can impact on braneworld cosmology.

\end{abstract}

\maketitle


\section{Introduction}\label{sec:intro}

Resonant states play a pivotal role in understanding many physical
systems from drums to molecules to black holes \cite{Nollert}.
They often take centre stage because they contain the key
observable features that are characteristic of the physical
system, irrespective of the initial conditions that excite them in
the first place. Braneworlds, it appears, are no
different~\cite{Seahra:2005wk,Seahra:2005us}. Even though a system
may be infinite and therefore unable to support a countable
collection of normal modes, there may exist discrete complex
frequencies which are nevertheless preferentially excited by
initial data -- so-called `quasi-normal modes' (QNMs). The real
parts of these frequencies are typically dependent on the length
scale of the system, while the imaginary part tells us the rate at
which energy is lost to infinity in each particular mode. A recent
investigation~\cite{Seahra:2005wk} has shown how the one-brane
Randall-Sundrum (RS) model \cite{Randall:1999vf} exhibits exactly
this behaviour: the continuous Kaluza-Klein (KK) spectrum of
gravitons has preferential masses with very short lifetimes.

Such resonant phenomena may have a significant effect in
braneworld cosmology.\footnote{See Ref.~\cite{Maartens:2003tw} for
a comprehensive review of all aspects of braneworld gravity.} As
is well known, the effective 4-dimensional formalism governing
perturbations on the brane is not closed; that is, there are
dynamical degrees of freedom on the brane whose evolution is
determined by non-local bulk effects. Any complete picture of
braneworld cosmology must include these degrees of freedom, but
this is notoriously difficult to implement in practice. Limiting
the discussion to one-brane models, we note that some progress has
been made in certain special cases
\cite{Langlois:2000ns,Gorbunov:2001ge,Koyama:2003sb,Koyama:2003yz,Kobayashi:2003cb},
by using various approximation schemes %
\cite{Koyama:2000cc,%
Langlois:2000iu,%
Barrow:2001pi,%
Leong:2001qm,%
Leong:2002hs,%
Battye:2003ks,%
Battye:2004qw,%
Koyama:2004cf,
Kobayashi:2004wy}%
, and direct numeric solution of the problem in 5 dimensions %
\cite{Hiramatsu:2003iz,Ichiki:2003hf,Ichiki:2004sx,Hiramatsu:2004aa}.

An alternative line of attack arises from realizing that the
behaviour of perturbations will be dominated by resonances of the
brane-bulk system, which are precisely the quasi-normal modes.
Given knowledge of the nature of these resonances, one should be
able to make reasonable predictions about the effect bulk gravity
waves have on the brane without knowing the details of how they
were generated.  This is especially useful since the correct
initial conditions for bulk gravity waves in such models are not
really known.

There are two aspects to the quasi-normal mode problem for
cosmological branes: the nature of the bulk, and the motion of the
brane. The latter is tricky because of the non-separability of the
equations of motion (the boundary condition in particular).
However, a static configuration is relatively straightforward, and
is a useful approximation to a cosmological model provided the
brane motion is `slow'.

A Friedmann-Lema\^\i tre-Robertson-Walker brane may exist in a
5-dimensional bulk provided that it is a piece of
Schwarzschild-anti-de Sitter (S-AdS$_5$) \cite{Maartens:2003tw}.
We shall focus on the case of a positive curvature S-AdS$_5$, with
an Einstein static brane around it, and choose the `inside' part
of the bulk (i.e., including the black hole), while enforcing
$\mathbb{Z}_2$ symmetry across the brane. In a previous
paper~\cite{Seahra:2005us} we have investigated tensor
perturbations when the brane is pure tension. In this situation we
found that gravity is `delocalized': gravity is sucked of the
brane by the presence of the black hole, removing the GR-like
zero-mode. Moreover, the presence of quasi-normal modes was
investigated and were shown to be quite heavily damped, but with a
mass-gap between the fundamental mode and the higher overtones.

The pure tension condition, while useful for comparison with RS
models, forces the brane to be located at the photon sphere of the
bulk, providing only a glimpse of the possible range of behaviour
relevant to cosmology. In particular, if the brane is moving it
will pass, in principle, from the white hole region through the
horizon and out to infinity (or maybe re-collapse). Hence we shall
consider all brane locations outside the horizon.

In this paper we show that the gravity waves are indeed dominated
by quasi-normal modes, the character of which depend strongly on
the brane location. Roughly speaking, the closer the brane to the
black hole horizon the higher the imaginary parts of the QNMs:
this corresponds to the stronger gravity of the black hole
draining gravitons of the brane more effectively. On the other
hand, for branes located far from the black hole, or on small
scales on the brane (provided the black hole mass is not too large
compared to the AdS length scale), the damping of the modes is so
small that the gravitons are effectively confined to the region
between the brane and the photon sphere of the black hole: in
effect, the quasi-normal modes become (approximate) normal modes,
and gravity waves behave akin to two-brane scenarios.  Thus, we
find what we are looking for: brane signals are effectively
`combed' in the frequency domain to be composed of a discrete
spectrum of \emph{very} lightly damped harmonic modes~-- sharp
evenly spaced spikes. Somewhat regardless of the initial data,
these spikes are the main feature of the signal on the brane.

The layout of the paper is as follows: In Section~\ref{sec:ES} we
discuss the Einstein static brane in S-AdS$_5$, and tensor
perturbations therein. In Section~\ref{sec:QNMS} we consider a
qualitative wave-mechanics analysis of the problem, before going
on to calculate QNMs for these models using a series solution for
the master wave equation. This techniques proves to be
computationally intensive for brane locations far from the
horizon, so in Section~\ref{sec:far brane} we use an alternative
approach for finding QNMs with small imaginary parts~-- which we
call \emph{quasi-bound states}. We then undergo a detailed study
of how these states behave in the three dimensional parameter
space which governs these models. We then discuss the utility of
our results in the time domain in Section~\ref{sec:IVP} using both
analytic Green's function and numerical methods.  We find the
existence of coherent states which would look like bouncing
gravitons to a brane observer; we also consider Hawking radiation,
and its distorted spectrum on the brane. Finally in
Section~\ref{sec:conc} we summarise our results and consider the
implications for cosmology.

\section{The Einstein-static braneworld}\label{sec:ES}

\subsection{Bulk geometry}

The bulk geometry of the brane universes considered in this paper
are given by the Schwarzschild-\ads{5} line element, which is
conventionally written as:
\begin{subequations}
\begin{eqnarray}
    \ds{5} & = & -f\,dT^2 + f^{-1}
    \,dR^2 + R^2 \, d\Omega_3^2, \\
    f & = & 1 -
    \frac{R_0^2}{R^2} +
    \frac{R^2}{\ell^2}.
\end{eqnarray}
\end{subequations}
Here, $R_0$ is related to the ADM mass of the black hole while
$\ell$ is related to the (negative) cosmological constant. It is
convenient to rewrite $f$ as
\begin{equation}
    f = \frac{(R^2 + R_\text{h}^2+\ell^2)(R^2 -
    R_\text{h}^2)}{R^2 \ell^2},
\end{equation}
where
\begin{equation}
    R^2_\text{h} = \frac{\ell^2}{2} \left( \sqrt{
    \frac{4R_0^2}{\ell^2} + 1} - 1 \right).
\end{equation}
When the solution is written in this way, it is obvious that there
is an event horizon at $R = \Rh$.  For simplicity, we will use
dimensionless $(t,r)$ coordinates defined by the substitutions:
\begin{eqnarray}\nonumber
    R & \rightarrow & r \times \Rh, \\ \mathcal{T}
    & \rightarrow & t \times \Rh, \\ \nonumber \ds{5} &
    \rightarrow & \ds{5} \times R_\text{h}^2.
\end{eqnarray}
In terms of these quantities, the bulk line element looks like
\begin{subequations}
\begin{eqnarray}
    \ds{5} & = & -f\,dt^2 + f^{-1}
    \,dr^2 + r^2 \, d\Omega_3^2, \\
    f(r) & = & \frac{(r^2 + \gamma^2 + 1)(r^2-1)}{\gamma^2 r^2}, \quad
    \gamma \equiv \frac{\ell}{\Rh}.
\end{eqnarray}
\end{subequations}
In this representation, the bulk geometry is completely
characterized by the ratio of the AdS length scale to the horizon
radius of the black hole $\gamma$ and the horizon is always at $r
= 1$. We shall call solutions with $\gamma \lesssim 1$ `big' black
holes and solutions with $\gamma \gtrsim 1$ `small' black holes.
In order to reintroduce dimensions to any quantity in the $(t,r)$
coordinates, one has to multiply by the appropriate power of the
horizon radius $\Rh$.

\subsection{Einstein-static branes and their matter content}

An Einstein-static brane is introduced by identifying some $r=\rb$
as the boundary $\Sigma_0$ of the 5-manifold and discarding the
portion with $r > \rb$.  The brane geometry is then simply that of
an Einstein static universe:
\begin{equation}
    \ds{4} = -d\tau^2 + r_\text{b}^2 \, d\Omega_3^2,
\end{equation}
where $\tau = f(\rb) t$ is the cosmic time.  The normal to
$\Sigma_0$ is selected to point \emph{away} from the brane and is
obtained by evaluating the following at $r=\rb$:
\begin{equation}
    n_a = -f^{-1/2} \di_a r.
\end{equation}
With this normal, one finds that the extrinsic curvature of
$\Sigma_0$ is
\begin{equation}\label{extrinsic 1}
    K_{ab} = h^c{}_b \nabla_c n_a = -\frac{1}{2f^{1/2}}
    \frac{df}{dr} u_a u_b - \frac{f^{1/2}}{r} q_{ab},
\end{equation}
where evaluation at $r = \rb$ is understood.  Here, we have
defined the objects
\begin{subequations}
\begin{eqnarray}
    u_a & = & -f^{1/2} \di_a t, \\
    h_{ab} & = & g_{ab} - n_a n_b, \\
    q_{ab} & = & h_{ab} + u_a u_b;
\end{eqnarray}
\end{subequations}
where $u^a$ is the future-directed normal to constant $t$
surfaces, $h_{ab}$ is the induced 4-metric on the brane, and
$q_{ab}$ is the spatial 3-metric.

The matter content of the brane is determined as follows: First,
assume a brane stress energy tensor of the perfect fluid form
\begin{equation}
    S_{ab} = \rhotot u_a u_b + \ptot q_{ab}.
\end{equation}
The fluid quantities $\rhotot$ and $\ptot$ refer to the {total}
brane density and pressure, {including} any contribution from the
brane tension $\sigma$.  Hence, these can be further decomposed as
\begin{equation}\label{matter decomposition}
    \rhotot = \rho_\text{m} + \kappa^{-1} \sigma, \quad \ptot =
    p_\text{m} - \kappa^{-1} \sigma,
\end{equation}
where $\kappa$ is a dimensionless 5-dimensional gravity-matter
coupling constant, while $\rho_\text{m}$ and $p_\text{m}$ refer to
the density and pressure of `ordinary' brane matter. Imposing
$\mathbb{Z}_2$ reflection symmetry across the brane yields (via
the standard junction conditions):
\begin{equation}
    \kappa S_{ab} = -2(K_{ab}-K h_{ab}),
\end{equation}
which is then equivalent to
\begin{equation}\label{extrinsic 2}
    K_{ab} = -\kappa \left( \tfrac{1}{3} \rhotot + \tfrac{1}{2}
    \ptot \right) u_a u_b - \tfrac{1}{6} \kappa \rhotot q_{ab}.
\end{equation}
Comparison of (\ref{extrinsic 1}) and (\ref{extrinsic 2}) yields
the following consistency conditions for an Einstein static brane:
\begin{subequations}\label{consistency}
\begin{eqnarray}\label{consistency 1}
   \left. \frac{1}{f^{1/2}} \frac{df}{dr} \right|_{r=\rb} & = & -2\kappa
    \left(\frac{\rhotot}{3}+\frac{\ptot}{2}\right), \\
    \label{consistency 2}
    \left.\frac{f^{1/2}}{r} \right|_{r=\rb} & = & \frac{\kappa\rhotot}{6}.
\end{eqnarray}
\end{subequations}
It is possible to derive these conditions from the effective
Friedman equation governing a general brane universe embedded in
S-\ads{5}:
\begin{equation}\label{friedman}
    \frac{\dot{a}^2}{a^2} = -\frac{f(a)}{a^2} + \frac{\kappa^2
    \rhotot^2}{36},
\end{equation}
where an overdot indicates differentiation with respect to cosmic
time.  A time derivative of (\ref{friedman}) yields the
Raychauduri equation for $\ddot a$, and setting $\dot a = \ddot a
= 0$ yields (\ref{consistency}) when the first law $d(\rhotot a^3)
= -\ptot d(a^3)$ is enforced and the identification $a=\rb$ is
made.

Equations (\ref{consistency}) comprise a system of 2 equations in
4 variables $\{\gamma,\rb,\rhotot,\ptot\}$.  Hence, there is a
2-parameter family of Einstein-static brane solutions.  For our
purposes it is useful to regard the size of the bulk black hole
$\gamma$ and the brane position $\rb$ as free parameters, while
the matter characteristics $\{\rhotot,\ptot\}$ are inferred from
(\ref{consistency}). It is easy to show that for the restrictions
$\gamma
> 0$ and $\rb > 1$, there is always a solution for
$\{\gamma,\rb,\rhotot,\ptot\}$. Furthermore, the effective
equation of state of the brane matter is
\begin{equation}
    w_\text{tot} \equiv \frac{\ptot}{\rhotot} = -\frac{3\rb^4-\gamma^2-1+2\gamma^2
    \rb^2}{3(\rb^2+\gamma^2+1)(\rb^2-1)} < -\frac{2}{3}.
\end{equation}
The inequality implies that a necessary condition for an
Einstein-static brane universe is $2\rhotot + 3\ptot < 0$.  In
other words, the total brane matter distribution must violate the
5-dimensional version of the strong energy condition.  This makes
sense since the brane is essentially a static shell of matter
around a black hole.  The only way to maintain such a
configuration is to construct the shell out of gravitationally
repulsive material, hence the violation of the strong energy
condition.  Note that does not necessarily hold for the
non-tension contribution to $\{\rhotot,\ptot\}$; i.e., it is
possible to have $\{\rho_\text{m},p_m\}$ consistent with the
energy conditions depending on the choice of $\sigma$.

Our principle concern is to understand how gravity waves change
with brane location. If the brane were allowed to move, these
conditions would no longer apply for a given brane position.

\subsection{Tensor perturbations}\label{sec:perturbations}

We now consider tensor perturbations of the bulk using the general
results of Kodama and Ishibashi \cite{Kodama:2003jz} specialized
to S-\ads{5}. Under such fluctuations:
\begin{equation}
    g_{ab} \rightarrow g_{ab} + \delta q_{ab}, \quad u^a \delta q_{ab} =
    n^a \delta q_{ab} = 0,
\end{equation}
where $\delta q_{ab}$ is the perturbed spatial 3-metric.

\begin{figure}
    \includegraphics{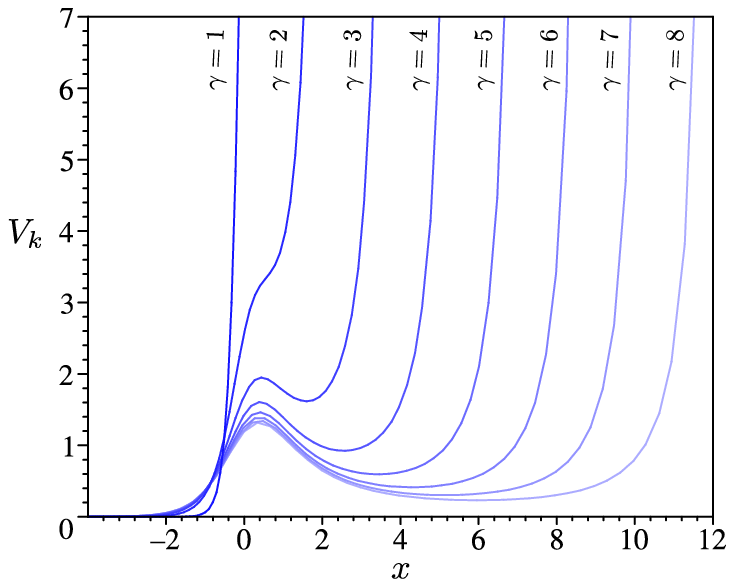}
    \caption{The potential in the master equation governing tensor
    perturbations in S-\ads{5} with no brane present and
    $L=1$}\label{fig:tensor potential}
\end{figure}

The SO(4) bulk symmetry allows a mode decomposition of $\delta
q_{ab}$ in terms of tensor harmonics on the unit 3-sphere $S^3$.
Let $i,j=1,2,3$ and $\nabla_i$ be the metric-compatible covariant
derivative on $S^3$. Then, tensor harmonics are defined by
\begin{equation}
    \Box \mathbb{T}^{(k)}_{ij} = -k^2 \mathbb{T}^{(k)}_{ij}, \quad
    \nabla^i \mathbb{T}^{(k)}_{ij} = 0 = \mathrm{Tr} \,
    \mathbb{T}^{(k)},
\end{equation}
where $\Box = \nabla^i \nabla_i$, and
\begin{equation}
    k^2 = L(L+2)-2, \quad L = 1,2,3,\ldots
\end{equation}
Now, if $\theta^i$ are suitable angular coordinates on $S^3$, the
harmonic decomposition of the metric perturbation reads:
\begin{equation}\label{expansion}
    \delta q_{ab} = \di_a \theta^i \di_b \theta^j
    \sum_k r^{1/2} \psi_k(t,r) \mathbb{T}^{(k)}_{ij}.
\end{equation}
When this is substituted into the linearized Einstein equations,
on finds that each of the $\psi_k$ satisfies a wave equation
\begin{equation}\label{master wave pde}
    -\frac{\di^2 \psi_k}{\di t^2} = - \frac{\di^2 \psi_k}{\di
    x^2} + V_k(r)\psi_k, \,\,\, x = x(r) \equiv \int\limits_r \frac{du}{f(u)},
\end{equation}
where $x$ is the tortoise coordinate and the potential is
\begin{equation}
    V_k(r) = f \left[ \frac{15}{4\gamma^2} +
    \frac{4 k^2 + 11}{4 r^2} + \frac{9 (\gamma^2+1)
    }{4\gamma^2 r^4} \right].
\end{equation}
The $L=1$ potential is shown in Figure \ref{fig:tensor potential}
for several values of $\gamma$.  We see that for large $\gamma$,
there is a barrier near $x = 0$ and an infinite wall at some
finite $x > 0$.  For small $\gamma$ the barrier is absent, but the
wall is still there.  The form of $V_k$ highlights one of the
curious properties of AdS space, namely $r(x) \rightarrow \infty$
at a finite value of $x$. This means that, unlike asymptotically
flat spacetimes, it is possible for gravity wave, electromagnetic,
or other types of signals to propagate from finite $r$ to
$r=\infty$ in a finite amount of coordinate time. The infinite
walls of the potential curves in Figure \ref{fig:tensor potential}
indicate the $x$ position of spatial infinity, and in the absence
of a brane one needs to specify boundary conditions there to have
a well-posed Cauchy problem for the evolution of $\psi_k$.

But we are interested in the case where there is a brane and the
asymptotic region is not there, so we need the boundary condition
on tensor perturbations at the brane location.  To find this, we
need to introduce 4-dimensional coordinates $y^\alpha$, where the
$\alpha,\beta\ldots$ indices run over the directions tangent
$\Sigma_0$. Then, 4-tensors are related to 5-tensors by
\begin{equation}
    T_{\alpha\beta} = \di_\alpha x^a \di_\beta x^b T_{ab}, \text{
    etc.}
\end{equation}
In the background geometry, the fact that the radial off-diagonal
metric elements $g_{r\alpha}$ are zero imply
\begin{equation}
    K_{\alpha\beta} = -\tfrac{1}{2} f^{1/2} \di_r h_{\alpha\beta}.
\end{equation}
Since tensor fluctuations have $n^a \delta g_{ab} = 0$, this
formula also holds after perturbation.\footnote{An alternative way
of saying this is that the vector $r^a = \frac{\di x^a}{\di r}$ is
parallel to $n^a$ both before and after perturbation.} Hence,
\begin{equation}
    \delta K_{\alpha\beta} = -\tfrac{1}{2} f^{1/2} \di_r \, \delta h_{\alpha\beta}
    \,\, \Rightarrow \,\, \frac{\kappa \rhotot}{3 f^{1/2}} \delta
    q_{\alpha\beta} = \di_r \, \delta q_{\alpha\beta}.
\end{equation}
Then, making use of (\ref{consistency 2}) to substitute for
$\kappa\rhotot f^{-1/2}$ and simplifying gives
\begin{equation}\label{boundary 1}
    0 = \di_r \left( r^{-2} \delta q_{\alpha\beta}
    \right)_{r=\rb}.
\end{equation}
It is worthwhile to contrast this with the boundary condition for
perturbations $\delta h_{\alpha\beta}$ in a Randall-Sundrum model
with scale factor $A(y)$ and in the Randall-Sundrum gauge: $\di_y
(A^{-2} \delta h_{\alpha\beta}) = 0$ on the brane.  Finally, we
use (\ref{boundary 1}) term-by-term in the expansion
(\ref{expansion}) to obtain the boundary condition satisfied by
the master variable:
\begin{equation}\label{master boundary cond}
    0 = \di_r \left( r^{-3/2} \psi_k \right)_{r=\rb},
\end{equation}
which is of a mixed Neumann-Dirichlet type.

\section{Quasi-normal gravity wave resonances}\label{sec:QNMS}

We have seen in Sec.~\ref{sec:perturbations} that tensor
perturbations of the Einstein static braneworld are governed by a
one-dimensional wave equation (\ref{master wave pde}) subject to a
boundary condition (\ref{master boundary cond}) at the brane. In
this section, we solve this equation via a series solution in
order to find the resonant modes of the potential, which are
analogous the the quasi-normal modes familiar from black hole
perturbation theory.

\subsection{Wave mechanics analysis}\label{sec:wave mechanics}

\begin{figure*}
    \begin{center}
    \includegraphics{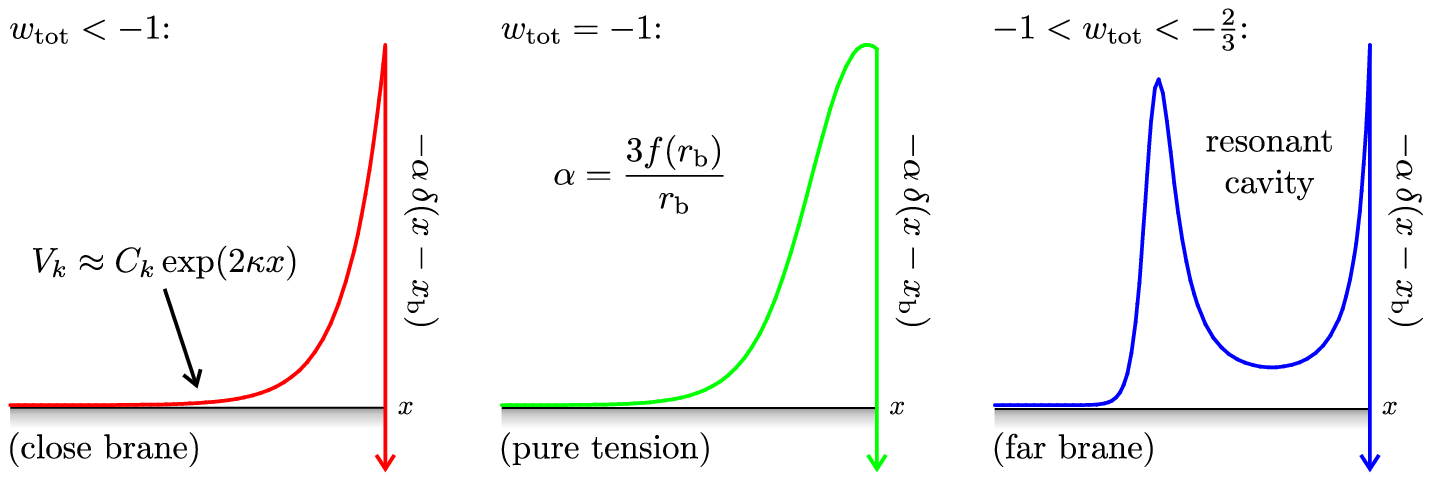}
    \end{center}
    \caption{Tensor gravity wave potentials for various
    Einstein-static braneworlds with a small bulk black hole.  We have taken
    $\gamma = 10$ and $L=1$ for these examples}\label{fig:ES potentials}
\end{figure*}%
If we substitute $\psi_k(t,r) = e^{i\omega t} \Psi_{k\omega}(x)$
into (\ref{master wave pde}), we obtain a time-independent
Schr\"odinger equation,
\begin{equation}\label{schrodinger eqn}
    \omega^2 \Psi_{k\omega} = -\Psi_{k\omega}'' + V_k(x) \Psi_{k\omega},
\end{equation}
with energy parameter $E = \omega^2$ and a non-standard boundary
condition
\begin{equation}\label{ode bc}
   0=[r^{-3/2}(x)\Psi_{k\omega}(x)]'_{r=\rb},
\end{equation}
at the position of the brane. However, this boundary condition
can be directly incorporated into the potential via an attractive
delta-function. In Fig.~\ref{fig:ES potentials}, we sketch the
resulting potential for several different brane positions in the
case of a small bulk black hole. There are two qualitatively
distinct regimes characterized by the effective equation of state
of the brane matter.  If this matter has a phantom equation of
state $w_\mathrm{tot} < -1$, we see that the brane is located in
between the bulk black hole's event horizon and photon sphere.  We
call this the `close brane' scenario.  If the equation of state is
more conventional $-1 < w_\mathrm{tot} < -\tfrac{2}{3}$, the brane
is located between the photon sphere and spatial infinity.
Correspondingly, this is known as the `far brane' scenario.  In
between the two possibilities is the pure tension case with
$w_\mathrm{tot} = -1$, when the brane is coincident with the
photon sphere.

The qualitative form of the potentials in Fig.~\ref{fig:ES
potentials} can give us valuable clues as to behaviour of gravity
waves in the ES braneworld.  We see that in all cases, the
potential vanishes like $e^{2\kappa x}$ as $x \rightarrow
-\infty$; i.e., at the bulk black hole horizon.\footnote{Though
not shown in Fig.~\ref{fig:ES potentials}, the vanishing of the
potential at the horizon is true for all values of $\gamma$.}
Here, $\kappa$ is the surface gravity of the black hole:
\begin{equation}\label{surface gravity}
    \kappa = \frac{1}{2} \frac{df}{dr} \Big|_{r=1} =
    \frac{2+\gamma^2}{\gamma^2}.
\end{equation}
As discussed in detail in Ref.~\cite{Seahra:2005us}, this means we
cannot find solutions for $\Psi_k$ with $\omega^2 > 0$ that are
localized near the brane. The vanishing of $V_k(1)$ means that the
horizon has zero reflectivity; that is, it is a perfect absorber
of gravity waves. One cannot construct a stable bound state under
such circumstances because there will always be energy leakage
into the bulk black hole.  This means that there is no
Randall-Sundrum like normalizable zero-mode for any
Einstein-static braneworld---the bulk black hole essentially
delocalizes the brane gravity.

If not stable bound states, what other types of resonances could
the potentials in Fig.~\ref{fig:ES potentials} support?  Well, the
existence of the attractive delta-function suggests that we may be
able to find metastable bound states.  There are modes that are
`almost bound' to the brane, but are eroded in time by the
tunnelling of the wave-function towards the bulk horizon.  We can
characterize such modes by a complex frequency with $\Imag\omega >
0$; i.e., they are exponentially damped in time.  Note that the
larger the imaginary part, the higher the damping and the shorter
the lifetime.  In this sense, they are the equivalent to the
familiar quasi-normal modes of black hole perturbation theory.
Indeed, metastable bound states are defined entirely analogously
to QNMs---namely, they are solutions of the wave equation
satisfying `purely outgoing' boundary conditions. In this context,
this means that $\psi_k$ satisfies both (\ref{master boundary
cond}) and
\begin{equation}
    \psi_k \sim e^{i\omega(t+x)}, \quad \text{as $x \rightarrow
    -\infty$.}
\end{equation}
Because we are imposing two boundary conditions on a single ODE,
it follows that the spectrum of QNM frequencies $\{\omega_i\}$ is
discrete. Now, consider the situation where a compact pulse of
gravity waves strikes the brane and scatters. A well-known result
it that the late-time behaviour of the scattered waveform will be
dominated by contributions from QNMs (so-called quasi-normal
ringing), independently of the initial data.  It is for this
reason that it is important to know the quasi-normal frequencies
of the brane: they represent the gravitational resonances of the
system and hence tend to dominate the gravity wave spectrum seen
on the brane.

As an important aside, we note that it is theoretically possible
to find QNM solution with $\Imag\omega_i<0$.  Such a solution will
be spatially normalizable because $\psi_k$ vanishes exponentially
as $x \rightarrow -\infty$.  It will also be exponentially growing
in time, and hence represents an instability of the system.  Such
an instability would not be unprecedented: It is well known that
the 4-dimensional Einstein-static universe is unstable to
homogeneous perturbations \cite{Barrow:2003ni}, and that a
Schwarzschild black hole enclosed in a finite cavity is unstable
to polar perturbations \cite{Gregory:2001bd} (assuming Dirichlet
boundary conditions).  This provides another motivation for
finding the QNMs of this system; i.e., to determine whether or not
it is stable against tensor perturbations.

Before we calculate the quasi-normal frequencies for the
Einstein-static braneworld, we comment on the qualitative features
of the various small black hole cases shown in Fig.~\ref{fig:ES
potentials}. The pure tension potential has been extensively
studied in Ref.~\cite{Seahra:2005us}, where it was found that the
QNM spectrum featured frequencies whose real and imaginary parts
were of the same order. How might we expect this to change for the
close and far brane cases?  For the far brane case, we see that
the peak of the potential and the brane form a kind of finite
potential well. It is plausible that gravity waves could be
temporarily trapped in this well, but would eventually tunnel out
to infinity. We expect a set of QNMs to be associated with this
effect, and their real parts should be roughly related to the
width of the resonant cavity $\lambda$. The trapping efficiency of
the cavity will be directly related to the lifetime of the QNMs,
and we will see in Sec.~\ref{sec:far brane} that the efficiency
can be very high indeed, resulting in quasi-normal mode
frequencies with tiny imaginary parts.

It is more difficult to guess the behaviour of the QNMs in the
close brane limit.  This should be considered as the very strong
gravity regime, where the delocalizing effects of the black hole
horizon are strongest.  So we may expect that the imaginary parts
of the QNMs to be much larger than the real parts.  In this case,
the potential is well-approximated by a simple exponential, which
allows for an analytic solution that we discuss in Appendix
\ref{sec:close brane}.

\begin{figure}
\begin{center}
    \includegraphics{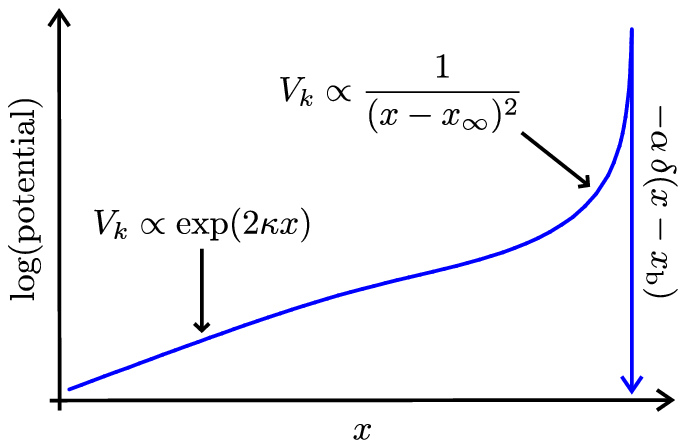}
\end{center}
\caption{The tensor potential for a large black hole, far brane ES
scenario with $(\gamma,L)=(0.5,1)$.  As in Fig.~\ref{fig:ES
potentials}, $\alpha = 3f(\rb)/\rb$ while $x_\infty =
x(\infty)$}\label{fig:large BH potential}
\end{figure}

Finally, we comment on the large black hole case.  As seen in
Fig.~\ref{fig:tensor potential}, there is no clearly defined
potential barrier for $\gamma \lesssim 2$, so the far brane
resonant cavity we discussed above will not be present for large
bulk black holes.  The actual far brane potential with $\gamma$
small is shown in Fig.~\ref{fig:large BH potential}.  Near the
brane, the potential is proportional to $1/(x-x_\infty)^2$, where
$x_\infty$ is the $x$-coordinate of spatial infinity. The
exponential fall-off of the potential as $x \rightarrow -\infty$
is unchanged from the small black hole case, and $V_k$ smoothly
interpolates between the two extremes.  The lack of a resonant
cavity implies that there will be no gravity wave trapping for far
branes, but the qualitative features of the near brane and pure
tension case should be the same as the small black hole scenario.

\subsection{Series solution}\label{sec:series}

Only so much progress can be made by looking at the potentials in
Figs.~\ref{fig:ES potentials} and \ref{fig:large BH potential}. So
we now turn our attention to an algorithm for actually calculation
the quasi-normal frequencies of the ES braneworld.  We begin by
analyzing the master wave equation (\ref{master wave pde}) in the
frequency domain, and transforming the radial coordinate to
\begin{equation}
    z = 1 - \frac{1}{r}.
\end{equation}
This coordinate maps the horizon at $r = 1$ onto $z=0$ and
$r=\infty$ onto $z=1$.  Dropping the $k$ index labelling the
different tensor harmonics, we write
\begin{equation}
    \psi(t,r) = e^{i\omega t} \phi_\omega(z),
\end{equation}
which converts (\ref{master wave pde}) into an ordinary
differential equation (ODE) of the standard form:
\begin{equation}
    \phi_\omega'' + P(z)\phi_\omega'+Q(z)\phi_\omega=0.
\end{equation}
where $P$ and $Q$ implicitly depend on $\omega$, $\gamma$ and $L$.
We will attempt to construct a series solution of the equation, so
we need to understand the pole structure of the continuations of
$P$ and $Q$ into the complex plane. One finds that $P(z)$ has
simple poles at:
\begin{equation}
    z = 0,2, \text{ and } 1\pm i(\gamma^2+1)^{-1/2}.
\end{equation}
The poles of $Q(z)$ are all of order 2 and are located at:
\begin{equation}
    z = 0,1,2, \text{ and } 1\pm i(\gamma^2+1)^{-1/2}.
\end{equation}
From this information, we see that the horizon $z = 0$ and spatial
infinity $z=1$ are regular singular points of the ODE about which
we can construct series solutions. Our problem involves discarding
the portion of the 5-manifold with $r > \rb$, so we should attempt
an expansion about $z = 0$.  Our solution will have a radius of
convergence greater than or equal to the distance to the closest
pole to $z = 0$; i.e., the solution will be valid for $z \in
[0,1]$.

Following the standard procedure, we substitute the ansatz
\begin{equation}
    \phi_\omega(z) = z^\varrho \sum_{m=0}^\infty a_m z^m
\end{equation}
into the ODE.  The coefficient of the lowest power of $z$ yields
an indicial equation with roots:
\begin{equation}
    \varrho = \pm \frac{i\gamma^2 \omega}{2(2+\gamma^2)}.
\end{equation}
Switching back to the $x$ coordinate, we see that the two
possibilities correspond to
\begin{equation}
    \phi_\omega \sim e^{\pm i \omega x} \text{ as } x \rightarrow -\infty.
\end{equation}
Hence, the two signs correspond to waves travelling into or away
from the black hole horizon.  For a QNM solution we need to choose
the former, which is the same as demanding that our perturbation
vanishes on the past black hole horizon. Once the sign of
$\varrho$ is fixed, it is a straightforward but tedious exercise
to obtain a ninth-order recurrence relation for the $a_m$
coefficients.

We have yet to impose the boundary condition at the brane, which
is
\begin{equation}\label{series boundary}
    \phi_\omega'(\zb) - \tfrac{3}{2} \rb \phi_\omega(\zb) = 0, \quad \zb \equiv 1 - 1/\rb.
\end{equation}
To deal with this, we define an $N$th order approximation to
$\phi_\omega$ via the partial sum:
\begin{equation}\label{partial sum}
    \phi_\omega^{(N)}(z) = z^\varrho \sum_{m=0}^N a_m z^m.
\end{equation}
Substitution of this into the boundary condition yields an
equation $\mathcal{P}_N=0$, where $\mathcal{P}_N$ is a polynomial
of order $2N$ in $i\omega$. Hence, only for a discrete set of $2N$
complex frequencies $\omega_n$ will our approximate solution
satisfy the brane boundary condition. Now, suppose that in the
limit $N \rightarrow \infty$ the roots of the \emph{polynomial
sequence} $\{\mathcal{P}_N\}_{N=1}^\infty$ converge to infinite
set of discrete frequencies. Then, such a set represents a
collection of resonant frequencies of tensor perturbations of the
bulk-brane system.

So, the problem of finding resonances for tensor perturbations of
Einstein-static braneworld amounts to finding roots of a sequence
of polynomials $\{\mathcal{P}_N\}$ that are stable in the $N
\rightarrow \infty$ limit.  The practical implementation of this
scheme is as follows: First, we fix all the input parameters of
the problem $\rb$, $L$, and $\gamma$.  Then, we begin with a low
order member of $\{\mathcal{P}_N\}$, of degree 30 in $\omega$ say,
and find all of its roots using Newton's method in the complex
plane. Then, we use these roots as initial guesses for Newton's
method applied to the polynomial of the next higher order.  If
Newton's method fails to converge after some number of iterations,
we discard the root. But if it does converge, we use the answer as
an initial guess for the next order polynomial, and repeat the
procedure.  We end up with a finite number of roots as a function
of $N$: $\omega_n = \omega_n(N)$. The final step is to reject any
of these roots that do not seem to be converging in the $N
\rightarrow \infty$ limit; i.e., roots for which the `convergence
factor'
\begin{equation}
    \Delta_n(N) = |\omega_n(N) - \omega_n(N-1)|,
\end{equation}
does not become small for $N$ large. At the end of the day, we are
left with a set of resonant frequencies for our choice of $\rb$,
$L$, and $\gamma$.

How big must we make $N$ in order to have accurate answers for
$\{\omega_n\}$? There is no entirely straightforward answer, but
note that for far branes with $\rb \gg 1$, we have $\zb
\rightarrow 1$. This means that we need to evaluate our series
solution near the singular point $z=1$ in order to enforce the
boundary condition (\ref{series boundary}). But we expect the
series to be poorly convergent near a singular point, so in order
to satisfy (\ref{series boundary}) accurately, $N$ must be very
large indeed. Hence, to find quasi-normal frequencies for $\rb \gg
1$, we need to retain more and more terms in our approximate
answer $\phi_\omega^{(N)}$. This is the principal practical
limitation of our method: the computational cost of computing
$\{\omega_n\}$ increases dramatically with increasing $\rb$.
Indeed, in Sec.~\ref{sec:far brane} we will need to resort to
other means to find the resonant modes for $\rb$ very large.

\subsection{Results}\label{sec:QNM results}

\begin{figure*}
\begin{center}
    \includegraphics{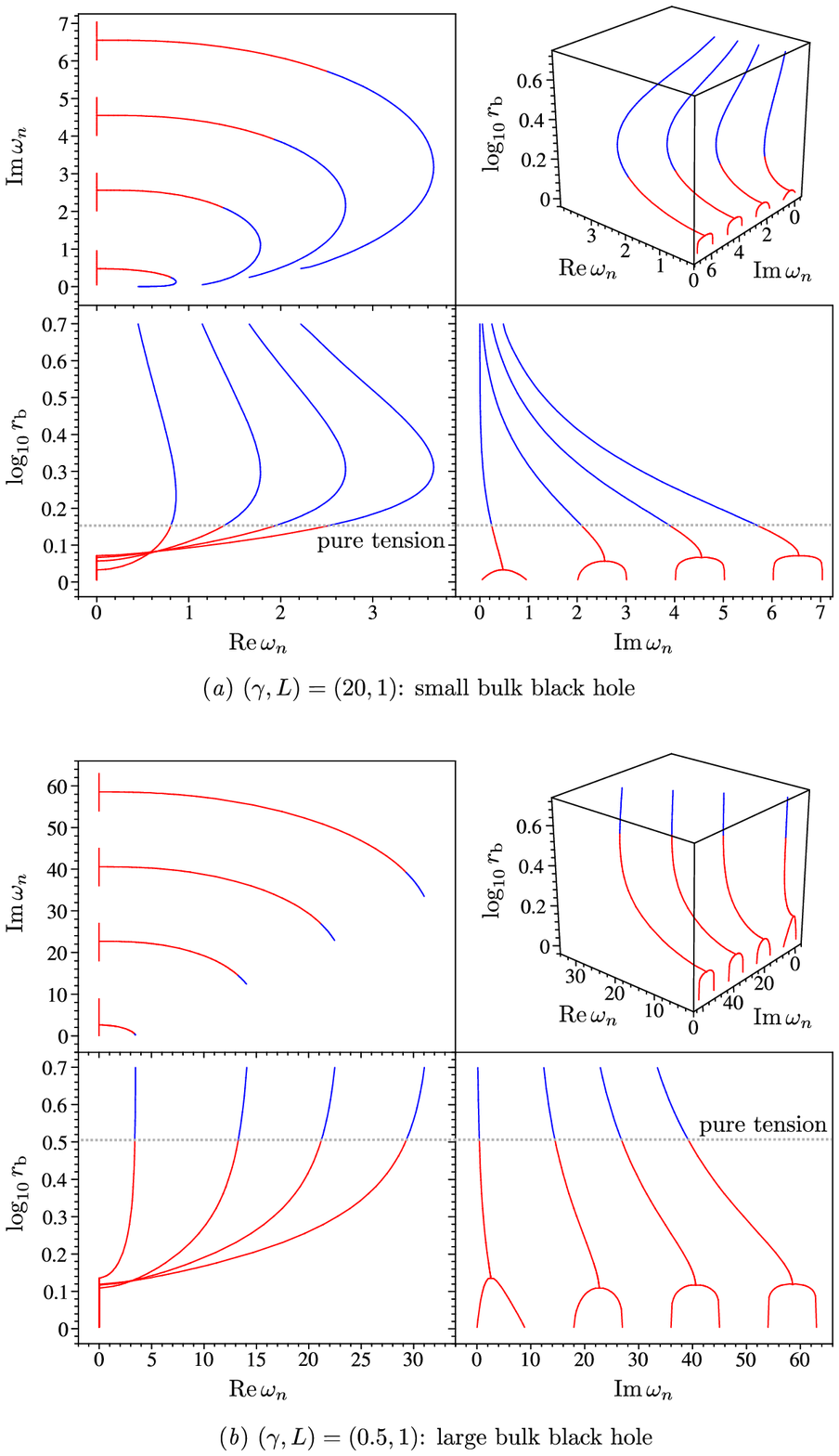}
\end{center}
\caption{QNM frequencies as a function of brane
position; close brane roots are colored red, while far brane
frequencies are blue}\label{fig:ribcage}
\end{figure*}

In Fig.~\ref{fig:ribcage}, we plot the several smallest $L=1$
quasi-normal frequencies of the brane for $\gamma = 20$ and
$\gamma = 1/2$.  As mentioned above, the maximum value of $\rb$ we
can consider is limiting by computing power, so we limit ourselves
to the range $\rb \in (1,5)$.  The roots of $\{\mathcal{P}_N\}$
have mirror symmetry about the imaginary axis, i.e., if $\omega_n$
is a quasi-normal frequency, then $-\omega_n^*$ is also a
solution. Hence, we only plot roots with $\Real\omega_n \ge 0$.

We highlight the key qualitative features:
\begin{itemize}

\item In both cases, the QNMs are purely imaginary in the
close brane regime, i.e., for $\rb \rightarrow 1$.  These are
overdamped, or \emph{evanescent} modes\footnote{This is the common
nomenclature in the theory of electromagnetic waveguides and other
fields.}, that show pure decay without oscillation.  Also, they
are evenly spaced in the limit, and the smallest root approaches
zero. In Appendix \ref{sec:close brane}, we show analytically that
\begin{equation}
    \lim_{\rb \rightarrow 1} \omega_n = i \kappa n, \quad n =
    0,1,2,3\ldots,
\end{equation}
where $\kappa$ is the surface gravity (\ref{surface gravity}). As
$\rb$ is increased, pairs of imaginary roots merge together and
then move off on a trajectory with nonzero real part.

\item Also in both cases, there is a tendency for the frequencies
to fan out and then migrate towards the real axis as $\rb$ is
increased further, past the pure tension threshold.

\item The two cases exhibit quite different behaviour in the far
brane regime. For a small black hole ($\gamma = 20$), the roots
move towards the real axis very quickly, ending up with
$\Real\omega_n$ much larger than $\Imag\omega_n$.  This means that
the several QNMs are lightly damped in this limit.  On the other
hand, the large black hole ($\gamma = 1/2$) has roots that drift
more slowly towards the real axis, resulting in moderately damped
modes.  The exception is the fundamental, defined as the QNM with
the smallest frequency, which becomes lightly damped in the far
brane limit.  Obviously, the reason for this discrepancy is that
small black hole potential (Fig.~\ref{fig:ES potentials}) has a
resonant cavity while the large black hole one
(Fig.~\ref{fig:large BH potential}) does not.

\end{itemize}
The extreme close and far brane behaviour will be discussed in
more detail in Secs.~\ref{sec:close brane} and \ref{sec:far
brane}, respectively.

But first, let us consider the implications of the pattern of
quasi-normal modes for actual gravity wave signals on the brane.
To do this, we fix $\gamma = 20$ (i.e. a small black hole) and
consider three separate brane positions corresponding to extreme
close, pure tension, and extreme far brane positions.  The four
smallest QNMs for each case are shown in Fig.~\ref{fig:QNM
comparison}.  We have plotted the imaginary parts of the modes on
a logarithmic scale because the damping varies by orders of
magnitude from case to case. The right axis shows the half-life,
defined as the time it takes for the modes amplitude to decrease
by a factor of $1/2$. The principal difference between the close
and pure tension cases is that the $\Real\omega_n = 0$ in the
former.  The half-lives of the mode are similar, and less than
about 10. However, the fundamental mode in the far brane case has
$t_{1/2} \sim 10^5$, which is extremely long compared to the other
cases. In fact, all of the far brane modes shown have very small
imaginary parts compared to the other cases.  This types of
behaviour are consistent with the speculation of
Sec.~\ref{sec:wave mechanics}---the resonant cavity is a efficient
gravity wave trap.
\begin{figure}
\begin{center}
    \includegraphics[width=\columnwidth]{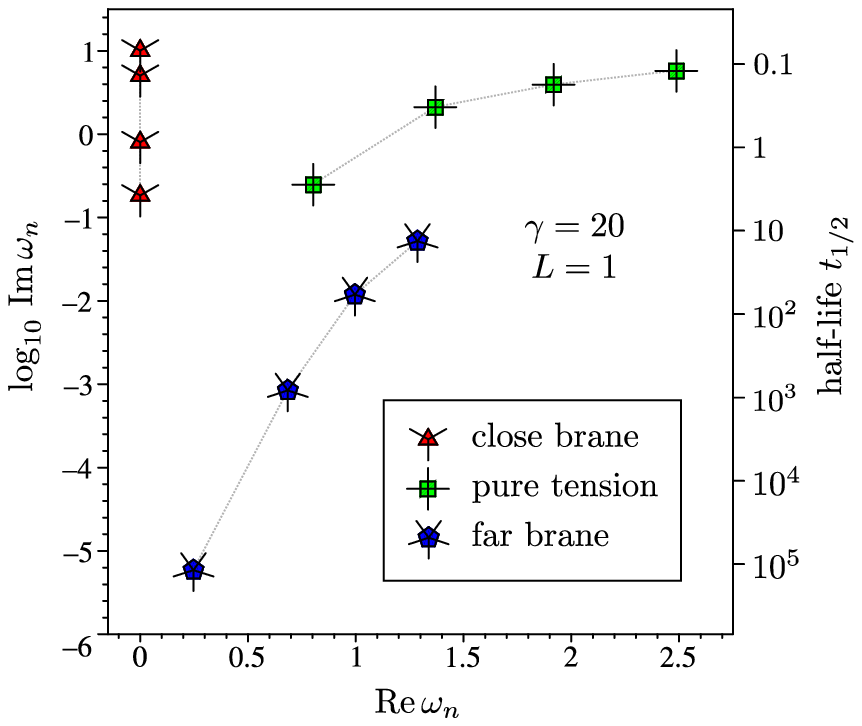}
\end{center}
\caption{QNM spectrum for close brane ($\rb = 1.05$), pure tension
($\rb = 1.42$), and far brane cases ($\rb = 10.00$) when $\gamma =
20$ and $L = 1$}\label{fig:QNM comparison}
\end{figure}

\begin{figure}
\begin{center}
    \includegraphics{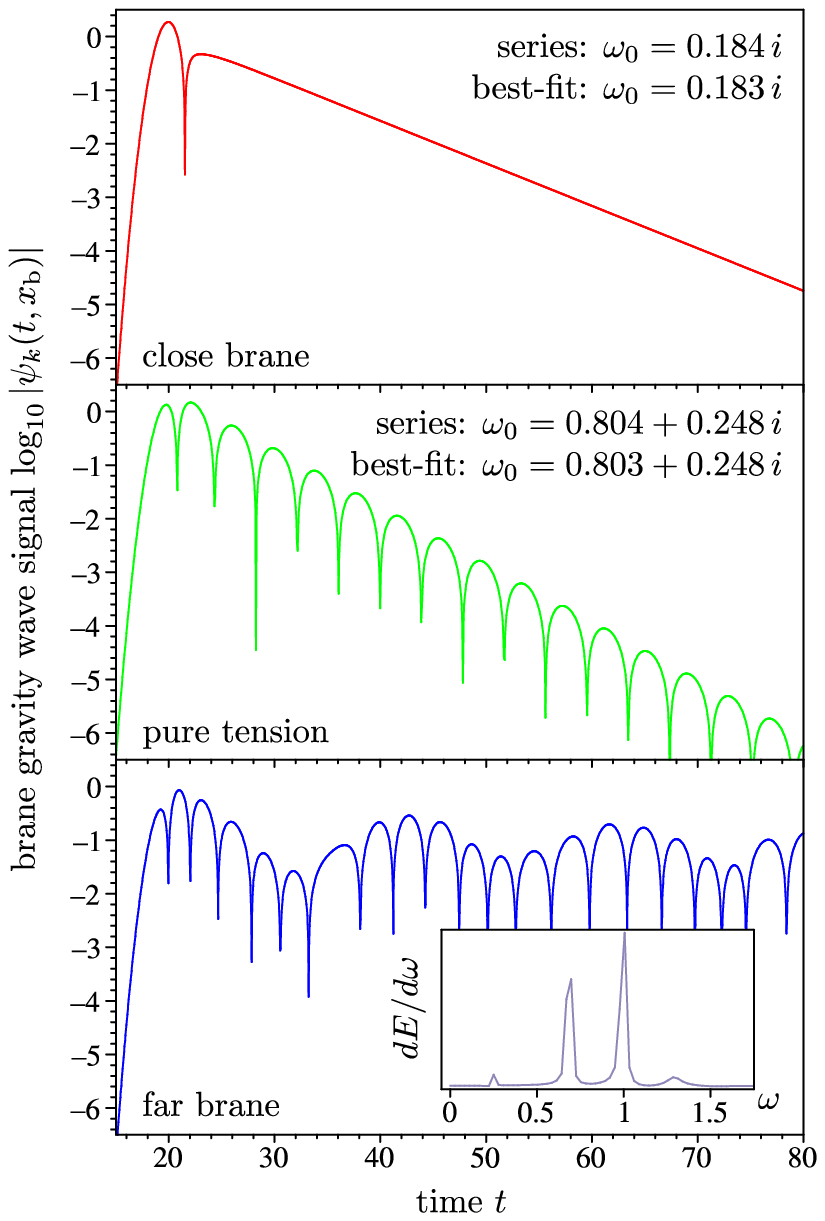}
\end{center}
\caption{Numeric brane gravity wave signals induced by a Gaussian
pulse striking the brane when $(\gamma,L) = (20,1)$ and $\rb =
1.05, 1.42, 10.00$ from top to bottom.  In all cases, the incident
pulse has a width of 2 and is centered about $x_0 = \xb - 20$ at
$t=0$. The `best-fit' values of $\omega$ in the top two panels
were found by matching a $\Real(A e^{i \omega t} )$ template to
the late-time signal, while the `series' value comes from
Sec.~\ref{sec:QNM results}. The inset for the last case shows the
Fourier transform of $\psi_k$ between $t = 25$ and
250}\label{fig:waveform comparison}
\end{figure}

We now perform `numeric scattering experiments' for each of the
three cases shown in Fig.~\ref{fig:QNM comparison}.  These consist
of numeric solutions of the PDE (\ref{master wave pde}) subject to
specified initial data.  In Fig.~\ref{fig:waveform comparison}, we
show the results of such integrations at the position of the
brane.  The initial data for each of the three cases was identical
Gaussian pulses incident on the brane at $t = 0$.  The resulting
signals reflect the characteristics of the QNM spectra seen in
Fig.~\ref{fig:QNM comparison}.  The close brane waveform is a
purely decaying exponential with no oscillations.  On the other
hand, the pure tension signal takes the form of a sinusoid with a
collapsing envelope.  In both of these cases, the fundamental
mode's lifetime is relatively large compared to the first
overtone, so we see signals dominated by one quasi-normal
frequency.  We can thus fit $\Real(Ae^{i\omega t})$ templates to
the signal to numerically determine the fundamental frequency
$\omega_0$.  These are the `best-fit' values shown in the upper
two panels.  Also shown are the frequencies determined from the
series solution shown in Fig.~\ref{fig:QNM comparison}, and the
agreement between the two methods is excellent.

The far brane signal in the lower panel is markedly different.  We
see in Fig.~\ref{fig:QNM comparison} that three of the far brane
QNMs have lifetimes longer that the horizontal axis of
Fig.~\ref{fig:waveform comparison}, and the fourth's half-life is
of the same order.  So we expect contributions from all of the
modes, which is exactly what is seen: a superposition of lightly
damped QNMs.  The signal is not adequately described by a single
complex exponential, so we cannot find $\omega_0$ by fitting.
Instead, we plot the absolute square of the fast Fourier transform
(FFT) of the signal between $t = 25$ and 250 in the inset.  The
power spectra is peaked at the real parts of the four QNMs,
showing that all of them contribute to the signal. What is
misleading about the Fourier transform, however, is the nature of
the peaks. We shall discuss this in detail in the next Section,
but for now let us note the following: The peaks are located at
the QNM frequencies; the signal near these goes like
$\exp{i(\varpi+i\Gamma/2)t}$, implying the power spectrum behaves
like a Lorentzian, $\propto1/[(\omega-\varpi)^2+(\Gamma/2)^2]$.
Thus we expect the widths of the peaks in Fig.~\ref{fig:waveform
comparison} to be of order of the square of the imaginary part of
the QNM: this is extremely narrow, and not easily resolvable
numerically.

To summarize this section: We have found that when the brane is
very close to the horizon, all the quasi-normal frequencies are
purely imaginary.  This makes the gravity wave signal a simple
decaying exponential, as in the top panel of
Fig.~\ref{fig:waveform comparison}.  For intermediate brane radii,
the real and imaginary parts of the QNMs have roughly the same
magnitude, resulting in a `classic' ringdown signal; i.e., the
monochromatic damped oscillations seen in the middle panel of
Fig.~\ref{fig:waveform comparison}.  The far brane behaviour
depends on the size of the bulk black hole.  If it is large, the
far brane behaviour is not much different than the pure tension
regime.  But if the black hole is small, the resonant cavity
between the brane and the photon sphere allows for a number of
lightly damped QNMs.  This results in a long-lived multi-frequency
brane signal, as seen in the bottom panel of
Fig.~\ref{fig:waveform comparison}.

\section{Quasi-bound states}\label{sec:far brane}

As mentioned above, the main drawback of our series method of
finding quasi-normal frequencies is that it becomes unfeasible as
$\rb \rightarrow \infty$.  This is indeed unfortunate, because as
Fig.~\ref{fig:waveform comparison} demonstrates, some of the most
interesting gravitational waveforms are associated with large
brane radii.  As seen in Figs.~\ref{fig:ribcage} and \ref{fig:QNM
comparison}, this is due to QNMs with very small imaginary parts.
We are naturally curious about the behaviour of these modes as
$\rb$ becomes larger and larger, which motivates us to find a
different method of locating the system's resonances when the
imaginary part is small. Our techniques will be based on collision
theory in ordinary quantum mechanics \cite{Taylor,LL} and numeric
solution of the wave equation in the frequency domain. Most
interesting effects are associated with the resonant cavity
between the brane and the bulk potential peak, so we will mostly
highlight the small black hole case in this section.  In
Sec.~\ref{sec:big BH}, we we indicate how the conclusions
translate into the big black hole scenario.

\subsection{The scattering matrix and the trapping
coefficient}\label{sec:trapping}

In the beginning of Sec.~\ref{sec:wave mechanics}, we wrote
$\psi_k = e^{i\omega t} \Psi_{k\omega}$ and hence transformed the
master wave equation into a time-independent Schr\"odinger
equation (\ref{schrodinger eqn}) with energy parameter $E =
\omega^2$ and boundary condition (\ref{ode bc}). Here, we consider
the (numeric) solutions of this ODE with $\omega$ real. As in
Sec.~\ref{sec:series}, we will simplify our notation by dropping
the wavenumber $k$ label from all quantities.

\begin{figure}
\begin{center}
    \includegraphics{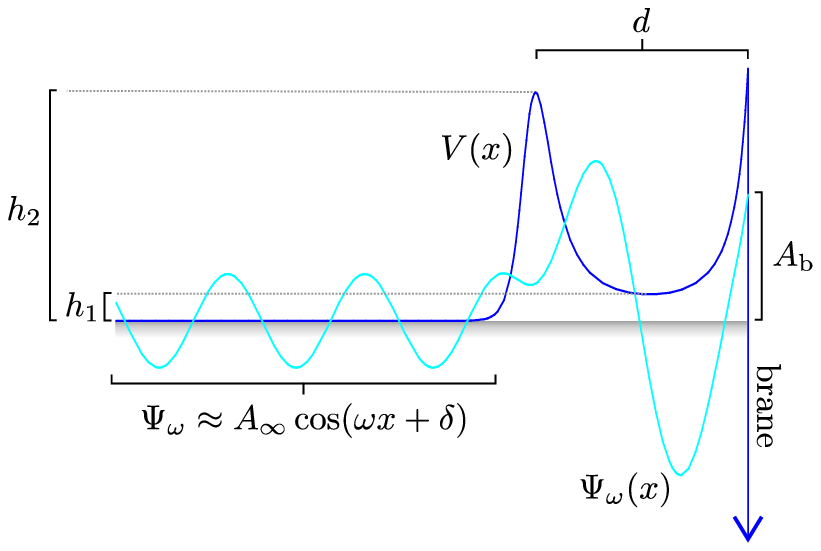}
\end{center}
\caption{Numeric solution of the master wave equation in the
frequency domain for a far brane, small black hole scenario. The
plot also shows the definitions of various quantities appearing in
the main text}\label{fig:ode solution}
\end{figure}

In Fig.~\ref{fig:ode solution}, we show the results of a typical
integration for a far ES brane, small black hole, and a given
value of $\omega$.  As could have been deduced from the potential,
the solution $\Psi_{\omega}$ satisfying the brane boundary
conditions takes the form of a simple sinusoid as $x \rightarrow
-\infty$. We can write this asymptotic solution as
\begin{equation}
    \Psi_{\omega} \approx \tfrac{1}{2} A_\infty e^{-i\delta} [e^{-i\omega
    x} + S(\omega)e^{i\omega x}],
\end{equation}
where $S(\omega) = e^{2i\delta}$ is the so-called \emph{scattering
matrix} and $\delta = \delta(\omega)$ is the \emph{scattering
phase shift}.\footnote{In our case, the scattering `matrix' is
clearly a scalar quantity. This is because we are really looking
at single-channel scattering, where there is no change of identity
of the graviton as it scatters off the brane. Multi-channel
scattering is common in nuclear physics, where the projectile and
target can turn into different particles after the collision,
which requires $S$ to be matrix-valued.} Presented in this way,
$\Psi_\omega$ is an explicit superposition of right and left
moving waves.

Now, we consider the continuation of $S$ to complex values of
$\omega$.  Suppose that the potential supports a QNM with
$\Imag\omega_n \ll 1$.  Then, we require that $S$ have a pole at
$\omega = \omega_n$ in order to ensure that $\Psi_\omega$ is a
purely outgoing solution.  For simplicity, we assume that this is
a simple pole.  Also note that the scattering matrix is unitary $S
S^* = 1$ when $\omega$ is real.  Under these circumstances, the
leading order Laurent expansion of $S$ about its pole must be of
the form \cite{Taylor}
\begin{equation}\label{laurent}
    S(\omega) \approx e^{2i\delta_\text{bg}} \frac{ \omega -
    \omega_n^*}{\omega - \omega_n},
\end{equation}
where $\delta_\text{bg}$ is the `background' phase shift, which is
an unspecified, slowly varying real function of $\omega$ for our
purposes.  If we now write the $n$th QNM as
\begin{equation}
    \omega_n = \varpi_n + \tfrac{1}{2} i\Gamma_n,
\end{equation}
we see that the total phase shift, with $\omega$ real and near to
$\varpi_n$, is of the form
\begin{equation}\label{phase shift}
    \delta(\omega) \approx \delta_\text{bg}(\omega) - \arcsin
    \frac{\Gamma_n}{\sqrt{4(\omega-\varpi_n)^2 +
    \Gamma_n^2}}.
\end{equation}
Because $\Gamma_n$ is small, the second `resonant' contribution
implies that $\delta(\omega)$ will vary rapidly about $\omega =
\varpi_n$.  Indeed, we expect that the phase will change by $\pi$
as $\omega$ is varied across the resonance.  The traditional way
of identifying a resonance is to calculate $\delta(\omega)$ for a
given potential and look for sharp features in $\sin^2
\delta(\omega)$.\footnote{Algorithms very similar to this
(so-called `resonance' methods) have been used to identify
lightly-damped QNMs of neutron stars \cite{Cha91,Kojima:1995nc}.}
Such lines will be centered about $\omega = \varpi_n$ and have
widths proportional to $\Gamma_n$, which means that long-lived
QNMs give rise to narrow features.

But is there a more physical quantity than $\sin^2\delta(\omega)$
to examine? Recall that QNMs are supposed to be metastable bound
states, so we intuitively expect that $\Psi_\omega$ to be
localized near the brane for $\omega \approx \varpi_n$.  We can
quantify this expectation by defining the \emph{trapping
coefficient}:
\begin{equation}
    \eta(\omega) = \frac{A_\text{b}}{A_\infty} ,
\end{equation}
where $A_\text{b}$ is the magnitude of $\Psi_\omega$ on the brane
(see Fig.~\ref{fig:ode solution}). Any state with $\eta(\omega)
\gg 1$ will correspond to the graviton being
more localized near the brane than at infinity; this will be a
`quasi-bound' state.
In fact, \emph{if} our potential supported a stable bound state
with $\Imag\omega_n = 0$, we would have $\eta(\omega_n) = \infty$.

Quantitatively, more can be said about $\eta(\omega)$ near
resonance by writing
\begin{equation}
    \Psi_\omega(x) = \Real [ e^{i\delta} \Psiout(x) ],
\end{equation}
where $\Psiout(x)$ is the purely outgoing solution
\begin{equation}
    \Psiout(x) =
    \begin{cases}
        e^{i\omega x}, & x = -\infty, \\
        Re^{i\theta}, & x = \xb.
    \end{cases}
\end{equation}
Here, $R$ and $\theta$ are assumed to be slowly-varying real
functions of $\omega$. This gives
\begin{equation}
    \xi(\omega) = \eta^2(\omega) = R^2(\omega) \sin^2
    [\delta(\omega)+\theta(\omega)-\pi/2].
\end{equation}
Now, suppose that the phases satisfy
\begin{equation}
\delta_\text{bg}(\omega) + \theta(\omega) \approx \pi/2 \text{ or
}3\pi/2, \quad \text{for } \omega \approx \varpi_n,
\end{equation}
then we have
\begin{equation}\label{lorentzian}
    \xi(\omega) \approx R^2(\omega) \frac{\Gamma^2_n}
    {4(\omega-\varpi_n)^2 + \Gamma_n^2}, \quad \text{for } \omega \approx \varpi_n.
\end{equation}
Under this assumption, we expect $\xi_\text{res}$ to be peaked
about the real part of each lightly damped QNM.  Furthermore, the
shape of the peak is Lorentzian (or Breit-Wigner) with a
half-width at half-maximum equal to the imaginary part of the
mode.  Of course, if $\delta_\text{bg} + \theta$ is not
approximately $\pi/2$ or $3\pi/2$ there is still a resonant
feature, but with a more complicated lineshape.

\begin{figure}
\begin{center}
    \includegraphics{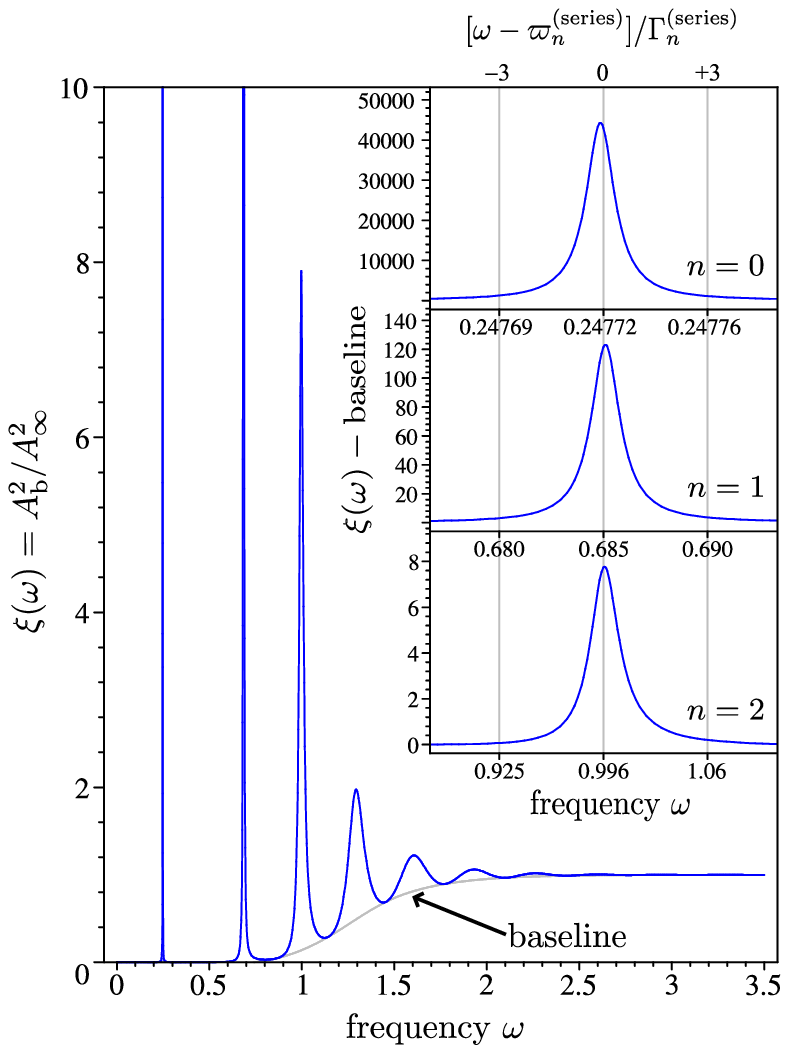}
\end{center}
\caption{The square of the trapping coefficient $\xi(\omega) =
\eta^2(\omega)$ in the $(\gamma,\rb,L)=(20,10,1)$ case. The insets
show the magnified views of the spectrum with a smooth
baseline---shown in the main figure---subtracted.  As described in
the text, we expect the baseline-subtracted peaks to have a
Lorentzian lineshape (\ref{lorentzian}) with $\varpi_n$ and
$\Gamma_n$ given by the series solution of Sec.~\ref{sec:QNM
results}.  We define the root-mean-square discrepancy between the
normalized peaks shown above $\xi_\text{t}$ and the series
prediction $\xi_\text{s}$, assuming a perfect Lorentzian, as
$\epsilon = [\int(\xi_\text{t}-\xi_\text{s})^2 \, d\omega]^{1/2} /
\int
\xi_\text{s}\,d\omega$.  This leads to $\epsilon = 0.011$,
$0.0073$, and $0.0072$, respectively, which are very small
discrepancies indeed}\label{fig:xi}
\end{figure}

The only way to constrain these `nuisance' phases is via direct
calculation. In Fig.~\ref{fig:xi}, we plot $\xi(\omega)$ for the
$(\gamma,\rb,L)=(20,10,1)$ case.  The spectrum features a number
of discrete peaks, and the positions of the first few peaks match
the real parts of the QNM frequencies calculated in
Sec.~\ref{sec:QNM results} for this case.  We can make this more
precise by subtracting a smooth baseline from $\xi(\omega)$ to
remove some of the non-resonant part of the spectrum; i.e.,
contributions from the variations of $\delta_\text{bg}$ and
$\theta$ with $\omega$. The insets of Fig.~\ref{fig:xi} show
magnifications of the baseline-subtracted spectrum about the
expected positions of the first three QNMs from the series
solution; and the frequency range of each inset has been set to be
20 times the imaginary part. The shapes of the three lines are
virtually identical and extremely well approximated by the
Lorentzian template (\ref{lorentzian}). Hence, this example
suggests that it is reasonable to take $\delta_\text{bg} + \theta
\approx \pi/2$ or $3\pi/2$ near an ES quasi-normal resonance.
Furthermore, we see that the \emph{height} of the peaks gives us
roughly $R(\omega)^2$, after the baseline is subtracted. We have
actually calculated the spectral function $\xi(\omega)$ for many
other cases, and found that the first few peaks in $\xi$ always
have a Lorentzian profile.

So, we have a straightforward algorithm for \emph{estimating}
lightly damped quasi-normal frequencies: First, find $\xi(\omega)$
from the numeric solution of the ODE (\ref{schrodinger eqn}).
Then, subtract a baseline and find all of the features in
$\xi_\text{res}$ with a Lorentzian profile.  The real and
imaginary parts of the frequencies are simply related to the
position and width of the those peaks.  We will call resonances
identified in this fashion \emph{quasi-bound states} (QBSs) for
two reasons: First, since they maximize the trapping coefficient
they are, in some sense, brane-localized modes with $A_\text{b}
\gg A_\infty$. Second, they all have $\Imag\omega_n \ll 1$; i.e.,
they are extremely long-lived excitations.  Of course, QBSs are
merely a sub-class of QNMs, but we will see below that they
dominate the phenomenology for far branes.

Finally, we should stress that this trapping coefficient method
really just estimates the QBS frequencies.  This is because the
procedure is prone to several uncertainties; including the
non-uniqueness of our choice of baseline, deviations from
Lorentzian profiles due to the running of $\delta_\text{bg}$, $R$
and $\theta$ with $\omega$, the validity of the Laurent expansion
(\ref{laurent}), etc. However, having said that, we find
remarkable agreement between this `trapping coefficient' method
and the series solution in the $(\gamma,\rb,L)=(20,10,1)$ case:
\begin{eqnarray}\nonumber
    \omega_0 & = &
    \begin{cases}
        0.247721+5.90 \times 10^{-6}\,i & \text{(series)} \\
        0.247720+5.88 \times 10^{-6}\,i & \text{(trapping)}
    \end{cases}, \\ \nonumber
    \omega_1 & = &
    \begin{cases}
        0.6852+8.45 \times 10^{-4}\,i & \text{(series)} \\
        0.6853+8.51 \times 10^{-4}\,i & \text{(trapping)}
    \end{cases}.
\end{eqnarray}
Such accuracy is more than sufficient for our purposes, so we now
proceed to use the `trapping method' to calculate the QBS
frequencies that the series solution has difficulties with.

\subsection{Resonant cavity quasi-bound state wavefunctions}

\begin{figure}
\includegraphics[width=\columnwidth]{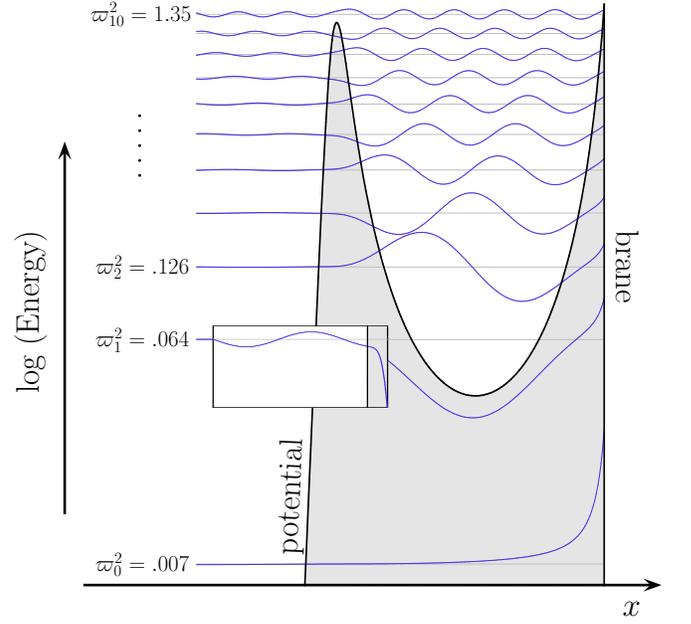}
\caption{Spatial wavefunctions $\Psi^n(x)$ of the first 11
quasi-bound states in the $(\gamma,\rb,L)=(20,250,1)$ case.
}\label{fig:eigenfunctions}
\end{figure}

Much intuition about the QBS solutions can be gained by examining
their spatial profiles $\Psi^n(x) = \Psi_{\varpi_n}(x)$.  We plot
these spatial profiles in Fig.~\ref{fig:eigenfunctions} for the
$(\gamma,\rb,L)=(20,250,1)$ case.  The mode with the smallest
frequencies is labelled with $n = 0$ and is the fundamental, while
the other modes are the `overtones'. The logarithm of the
potential $V(x)$ is shown in the background.
The vertical offset of each $\Psi^n$ corresponds to the
    energy parameter $E_n = \varpi_n^2$ defining the solution.  We see
    that the overtone energies are greater than the potential
    within the resonant cavity, while the fundamental's energy
    only becomes greater than $V$ in the asymptotic region.
    (Referring to Fig.~\ref{fig:ode solution}, this can be
    expressed as $\varpi_0^2 < h_1$, while $\varpi_n^2 > h_1$ for $n =
    1,2\ldots$)  As in ordinary quantum mechanics,
    the wavefunctions have concave curvature ($\Psi''/\Psi < 0$)
    when their energy is greater than the potential, and convex
    curvature ($\Psi''/\Psi > 0$) otherwise.  Hence, the overtone profiles
    oscillate in the resonant cavity while the fundamental shows
    monotonic decay.

    The integer $n$ actually counts the number of nodes~-- places
    where the
    wavefunction crosses zero
    between the brane and the
    peak in $V$.  This is also the number of local
    extrema---or `loops'--- in the wavefunction in this region.
    That is, $\Psi^1$ has one minima in the resonant
    cavity, $\Psi^2$ has one minima and one maxima, etc.  And the
    fundamental has none.   This type of behaviour is strongly
    reminiscent of energy eigenfunctions in quantum mechanics, as
    in the radial eigenfunctions of the hydrogen atom, for
    example.

         The distribution of energies causes the $n=0$ mode to
    be the only wavefunction sharply localized near the brane.
    Conversely, the overtone wavefunctions are dispersed throughout
    the resonant cavity (and beyond for high-energy modes).  This
    type of behaviour generalizes to all of the other cases we
    have looked at.

    The key point here is that the
     `quasi-bound state' moniker is well deserved for the
    lowest-lying modes, whose amplitude at infinity $A_\infty$ is
    vanishingly small.  The inset shows a magnified view of the $n
    = 1$ wavefunction in the asymptotic region to emphasize that
    it is actually sinusoidal and nonzero.  The shape of the
    overtone wavefunctions suggests a tunnelling process where
    `most' of the graviton is confined to the right of the barrier,
    but a small portion leaks through to
    infinity.  This leakage is more pronounced for higher energy
    modes, corresponding to larger imaginary parts~-- so wider Lorentzian
    profiles~-- for these frequencies.

Now that we have a fuller understanding of the trapping method and
the quasi-bound states which it finds, we turn to a deeper
investigation of the three dimensional parameter space of the
model. We shall look at varying $\rb$, $L$, and $\gamma$ in turn.


\subsection{Far brane limit: recovering the
zero-mode and evenly-spaced overtones}\label{sec:zero mode}

\begin{figure}
\begin{center}
    \includegraphics{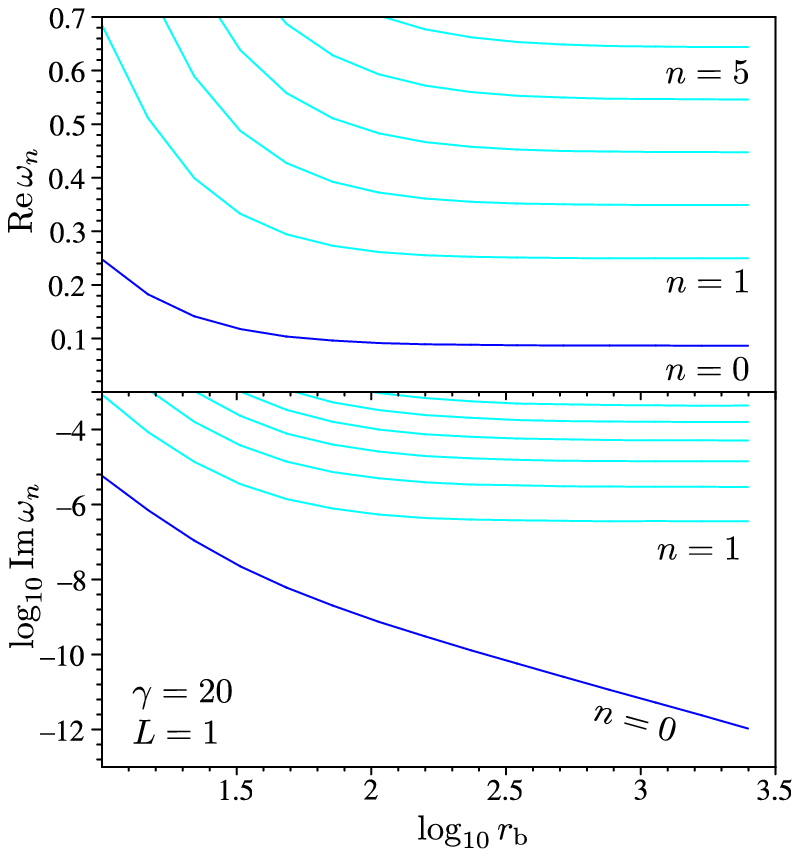}
\end{center}
\caption{Quasi-bound state frequencies as a function of brane
position. Note that the fundamental frequency $\omega_0$ tends to
a real value as $\rb \rightarrow \infty$, while the imaginary
parts of the overtone frequencies are asymptotically constant and
non-zero.}\label{fig:far brane qnm}
\end{figure}

The trapping method allows us to further investigate the trends in
the QNM frequencies shown in Fig.~\ref{fig:ribcage} to larger
values of $\rb$ in the small black hole case ($\gamma$ large).  In
Fig.~\ref{fig:far brane qnm}, we show the real and imaginary parts
of the quasi-normal frequencies when $(\gamma,L) = (20,1)$ as a
function of $\rb$. The most striking feature of this plot is the
disparate behaviour of the fundamental and overtone modes.  While
the imaginary parts of the overtones approach constant values as
$\rb \rightarrow \infty$, $\Imag\omega_0$ decays to zero as a
power law.  Hence, for infinitely large branes, we recover a
single stable graviton mode confined to the brane.

\begin{figure}
\begin{center}
    \includegraphics{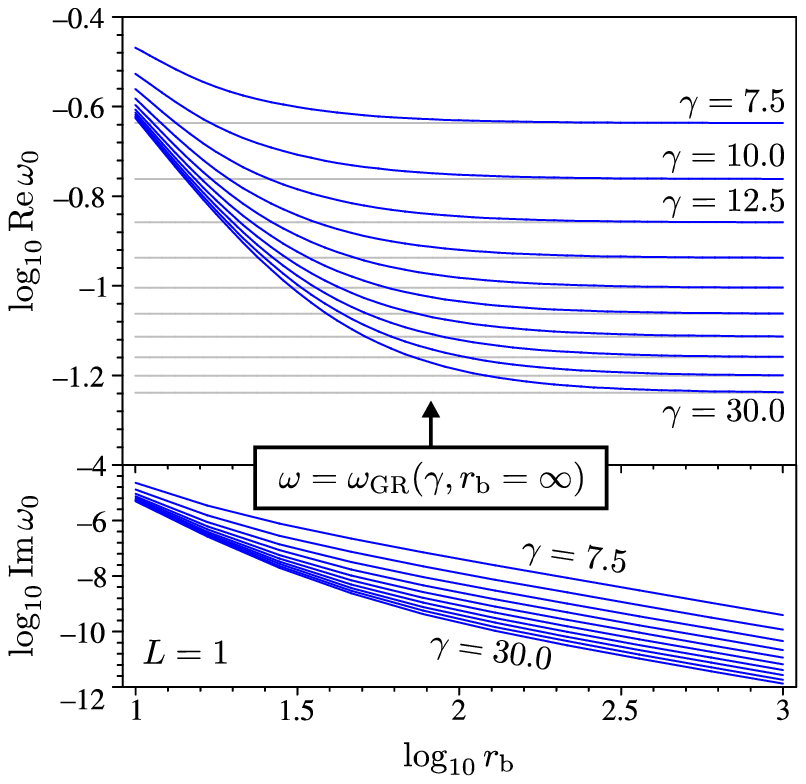}
\end{center}
\caption{Frequency of the fundamental QBS for varying $\gamma$ and
$\rb$.  The horizontal lines in the top panel indicate the GR
prediction for the frequency of tensor perturbations in an ES
universe as $\rb \rightarrow \infty$. Clearly, $\omega_0$
approaches $\omegaGR$ in the limit, suggesting that we recover a
RS-like zero-mode for infinitely large
branes.}\label{fig:fundamental}
\end{figure}

But is this infinite lifetime excitation the ES generalization of
the RS zero-mode? In 4 dimensions, tensor perturbations of an
Einstein static universe would oscillate with frequency
$\sqrt{L(L+2)}/\rb$ according to comoving observers
\cite{Barrow:2003ni}. Taking into account the redshift factor
between the bulk coordinates and a comoving brane observer, this
translates into a bulk frequency
\begin{equation}
    \omegaGR (\gamma,\rb) = \frac{\sqrt{\fb L (L+2)}}{\rb}.
\end{equation}
In Fig.~\ref{fig:fundamental}, we plot the fundamental QBS
frequency for a range of $(\gamma,\rb)$ parameters.  Also shown as
horizontal lines in the top panel is the GR frequency in the $\rb
\rightarrow \infty$ limit.  We see very clearly that for all
$\gamma$, $\omega_0$ approached $\omegaGR$ asymptotically.  Hence,
not only do we have a mode with infinite lifetime in the large
brane limit, the frequency of this mode matches what one might
expect from ordinary 4-dimensional theory.  These are the
essential characteristics of the standard zero-mode in the RS
scenario, so we conclude that as $\rb \rightarrow \infty$ we
recover the zero-mode in the ES model.

This is a very sensible result.  We saw above that the $n = 0$
wavefunction is distinguished from the others in several important
ways, including being the only mode to be truly localized on the
brane and nowhere else.  So it is not surprising to find that its
frequency shows a different dependence on $\rb$. Furthermore, in
Ref.~\cite{Seahra:2005us} we argued that the very existence of the
bulk black hole was responsible for the delocalization of the
zero-mode in the ES scenario.  Here, we see that as the effects of
the black hole at the brane position get smaller and smaller as
$\rb$ is increased, we `re-localize' brane gravity.  In other
words, the gravitational field of the bulk black hole strips the
zero-mode off the brane, but when the brane is far away this
delocalization becomes more and more inefficient.

Some of the features seen in Figs.~\ref{fig:far brane qnm} and
\ref{fig:fundamental} can be understood using WKB-approximation
techniques.  In quantum mechanics, decay constants for metastable
bound states can be approximated by the WKB tunnelling amplitude
across the potential barrier responsible confining the
wavefunction.  The classic example of this kind of calculation is
Gamov's description of alpha-decay using elementary wave
mechanics.  As seen in Fig.~\ref{fig:eigenfunctions}, the energy
of the fundamental mode lies beneath the resonant cavity; i.e.,
$\varpi_0^2 < h_1$. Hence, we can view its decay as a tunnelling
process from the delta-function at the brane position.  This gives
rise to
\begin{equation}\label{WKB imag}
    \Imag\omega_0 \sim \exp\left[ - \int\limits_{x_0}^{\xb}  \sqrt{
    V(x)-\varpi_0^2} \, dx \right].
\end{equation}
Here, $x_0$ is the `classical' turning-point such that $\varpi_0^2
= V(x_0)$.  Recall from Figs.~\ref{fig:tensor potential} and
\ref{fig:large BH potential} that the potential diverges at
spatial infinity.  The implies that as $\rb \rightarrow \infty$,
the rightmost portion of the integrand diverges like $(x_\infty -
\xb)^{-1}$, which in turn gives $\Imag\omega_0 \rightarrow 0$.
Furthermore, it is not difficult to show that
\begin{equation}
    \Imag\omega_0 \propto r_\text{b}^{-\sqrt{15}/2}, \quad
    \text{as $\rb \rightarrow \infty$}.
\end{equation}
This is exactly the asymptotic behaviour seen in
Fig.~\ref{fig:fundamental}: $\log_{10}\Imag\omega_0$ versus
$\log_{10}\rb$ is a straight line for large brane radii whose
slope is independent of $\gamma$.  We expect this behaviour to
generalize to other choices of $L$.

Turning our attention to the overtone modes, we note that their
energies satisfy $h_1 < \varpi_n^2 < h_2$, so there are three
turning points $x_0 < x_1 < x_2$.  The potential barrier that is
responsible for the quasi-confinement is the peak in $V(x)$ shown
in Fig.~\ref{fig:eigenfunctions}.  This peak does not become
infinitely large in the limit, which explains why the overtones
have finite damping asymptotically.  In this case, WKB techniques
can tell us something about the real parts of the QBS frequencies.
Because the overtone energies are greater than the potential
inside the resonant cavity, we can use the Bohr-Sommerfeld
quantization rule \cite{LL} to write:
\begin{equation}
    \int\limits_{x_1}^{x_2} dx \, \sqrt{\varpi_n^2-V(x)} =
    (n+\tfrac{1}{2})\pi, \quad n \ge 1.
\end{equation}
If we approximate the potential in the cavity by its average
value, we obtain:
\begin{eqnarray}\nonumber
    \varpi_n & \approx & \sqrt{ \bar{V} + \left[ \frac{(n+\tfrac{1}{2})\pi}
    {\Delta x} \right]^2} \\ & \rightarrow & \frac{(n+\tfrac{1}{2})\pi}
    {\Delta x} \text{ for $\bar{V} \ll \frac{n^2}{(\Delta x)^2}$},
\end{eqnarray}
where $\Delta x = x_2 - x_1$.  The last approximation holds
reasonably well for the higher modes in the current problem.
Finally, as $\rb \rightarrow \infty$ we have $\Delta x \approx d$,
where $d$ is the distance between the brane and the potential peak
(\emph{cf.}~Fig.~\ref{fig:ode solution}).  In the limit $d$
becomes constant, which explains why $\Real\omega_n$ becomes
asymptotically constant and evenly spaced for the overtones in
Fig.~\ref{fig:far brane qnm}.

\subsection{Small scale limit: the quasi-bound states become bound}

So far we have concentrated on perturbations with $L=1$; now
consider perturbations with $L$ not necessarily equal to unity.
The value of $L$ controls the wavelength of the gravity waves
tangent to the brane, so when we increase $L$ we are examining
smaller and smaller scales. What kind of modes will exist in this
case?

\begin{figure}
\includegraphics[width=0.9\columnwidth]{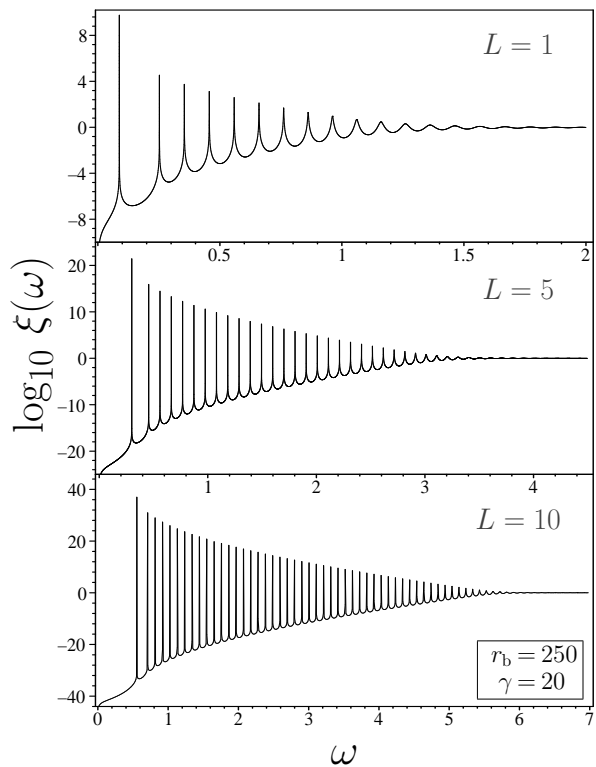}
\caption{Variation of the trapping coefficient with $L$.}
\label{fig:eta varying L}
\end{figure}

In Fig.~\ref{fig:eta varying L}, we plot the logarithm of the
trapping coefficient $\xi(\omega)$ for the $(\gamma,\rb) =
(20,250)$ case, but now letting $L$ increase from 1 through 5 to
10. The principle effect of increasing $L$ is to increase the
number and sharpness of the spikes~-- implying more quasi-bound
states with much lighter damping. The WKB approximation can again
be used to understand this behaviour.

\begin{figure}
\includegraphics{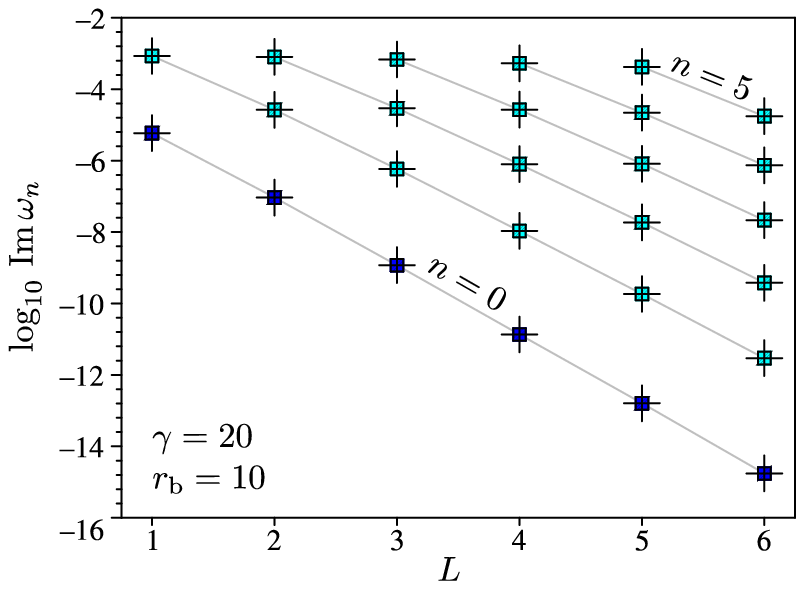}
\caption{The variation of the imaginary parts of QBS frequencies
with $L$.}\label{fig:varying L}
\end{figure}

Returning to (\ref{WKB imag}), we take the $L\rightarrow
\infty$ limit instead of $\rb \rightarrow \infty$. The height of
the peak in the potential scales like
\begin{equation}
    h_2 \approx \frac{(\gamma^2+2)^2}{4\gamma^2(\gamma^2+1)} L^2,
\end{equation}
so its contribution dominates the integral. The same holds if we
apply the formula to the overtones by switching $\varpi_0$ to
$\varpi_n$ and $\xb$ to $x_1$.  In any case, this leads to
\begin{equation}
    \Imag\omega_n \propto e^{-aL}, \quad \text{as } L
    \rightarrow \infty,
\end{equation}
where $a$ is a constant depending on $\gamma$ and $\rb$. This
exponential decay in the widths of the spikes is shown in
Fig.~\ref{fig:varying L}, where we plot the imaginary parts of the
QBS frequencies for various scales.

Physically, this is intriguing: As the spatial scale of
perturbations on the brane decreases, their bulk angular momentum
around the black hole increases.  This causes the centrifugal
barrier between the brane and the event horizon to grow taller and
taller, which in turn makes it difficult for gravity waves to
tunnel away from the brane into the black hole.  The net result is
that the confinement of the QBSs in the resonant cavity becomes
100\% efficient.  Effectively, the quasi-normal modes become
normal modes when their imaginary parts disappear; the quasi-bound
states become bound when $A_\infty$ goes to zero.  This is like a
two-brane model where the potential barrier takes the role of the
shadow brane.  We will discuss the implications for cosmology
below.

How many QBS states are there in the large $L$ limit?  Above we
saw that high overtones have $\omega_n \sim \pi(n
+\tfrac{1}{2})/d$.  The highest QBS mode should have have an
energy comparable to the height of the potential peak; i.e.,
$\omega_N^2 \sim h_2$.  Finally, we know that the height of the
peak scales as $L(L+2)$, which yields that the number of
quasi-bound states is proportional to $L$ as $L \rightarrow
\infty$.

\subsection{The big black hole case}\label{sec:big BH}

To this point, we have only applied the trapping coefficient
method to small black hole cases featuring a resonant cavity.  But
under certain circumstances, we can also use the procedure when
$\gamma$ is small.

The square of the trapping coefficient for range of relatively
small values of $\gamma$ and $\rb$ are shown in Fig.~\ref{fig:big
BH}. All the spectra show one peak indicating the existence of
only one QBS solution.  This is because there is no resonant
cavity when $\gamma$ is small (see Fig.~\ref{fig:large BH
potential}), hence there are no overtones.  The general trend is
for the peak position to show little variation with $\rb$, and to
decrease with increasing $\gamma$.  On the other hand, the width
of the feature decreases as either $\gamma$ or $\rb$ is increased.
\begin{figure}
\includegraphics{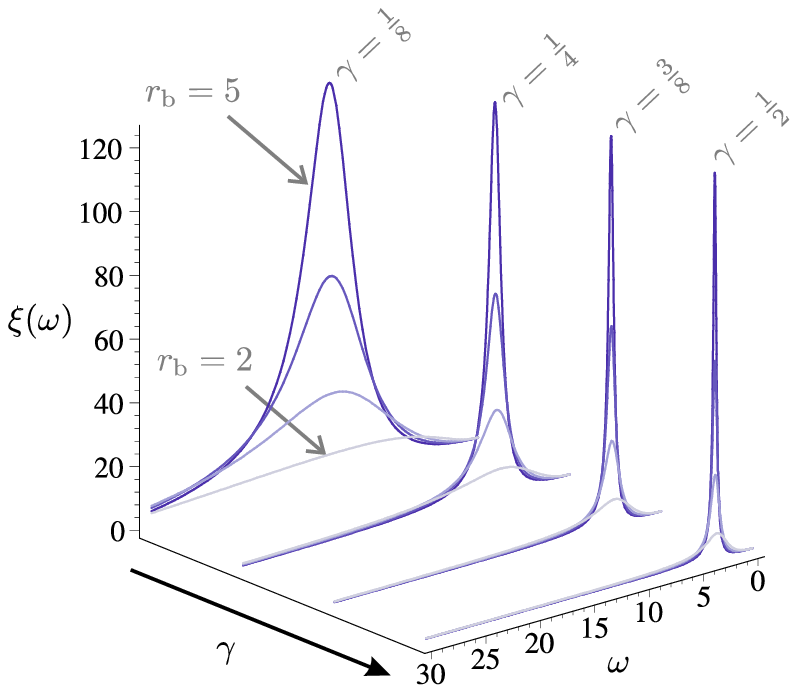}
\caption{$\xi(\omega)$ for various $L = 1$ big black hole
scenarios}\label{fig:big BH}
\end{figure}

The rather broad nature of the peaks for small $(\gamma,\rb)$
means that they are not well approximated by Lorentzian profiles,
which suggests no modes with tiny imaginary parts.  This is
consistent with the series results of Sec.~\ref{sec:QNM results},
where small imaginary parts were only achieved for large brane
radii. When $\rb$ is large, the fundamental-mode WKB analysis of
Sec.~\ref{sec:zero mode} is applicable to this case, implying that
the imaginary part of the sole QBS frequency goes to zero for
large branes.  We have confirmed this numerically, and also
confirmed that the real parts approach $\omegaGR$.  So, we also
recover the GR zero-mode as $\rb \rightarrow \infty$ in the big
black hole case.

As in the small black hole case, the potential barrier between the
brane and the horizon scales as $L(L+2)$ for large $L$.  Hence,
the conversion of quasi-bound into bound states as $L \rightarrow
\infty$ also occurs for small $\gamma$, the chief difference being
that there is only the fundamental mode to covert.

The main difference between the small and large black hole cases,
then, is that in the former there are quasi-bound state overtones,
while in the latter there are not.  However, there is always a
fundamental $n=0$ QBS, and the limiting behaviour of that mode as
$\rb$ or $L$ becomes large is the same irrespective of $\gamma$.

\section{The initial value problem}\label{sec:IVP}

In Sec.~\ref{sec:QNM results}, we solved the wave equation in the
time domain to illustrate the effects of the QNMs on actual brane
signals.  Here, we examine this problem in more detail by
considering several different classes of initial data.  We will
demonstrate that the trapping coefficient $\eta$ is useful for
more than just finding QBSs; it actually plays a prominent role in
determining the power spectrum of the brane signal.  We also
consider the scattering of wave packets in scenarios featuring a
resonant cavity, and observe coherent `bouncing' behaviour.
Finally, we allow for `initial data' representing the situation
where the brane is in thermal equilibrium with the semi-classical
Hawking flux from the black hole.

\subsection{Green's function analysis}\label{sec:greens}

Suppose that at $t = 0$, we specify the both the value and first
time derivative of $\psi$.  Then, we can construct a formal
Green's function solution for the signal on the brane $\psib (t) =
\psi(t,\rb)$ by following the standard technique from black hole
perturbation theory~\cite{Nollert,Andersson:1996cm}.  This
involves applying the Laplace transform to the master wave
equation to convert the initial data into a source term, obtaining
the Green's function solution, and then performing the reverse
transform to return to the time domain.  If we write the resulting
solution as an integral in Fourier space, we obtain
\begin{equation}
    \psi(t,x) = \frac{1}{2\pi} \int d\omega \, dx' e^{i\omega
    t} G_\omega(x,x') \mathcal{I}(x',\omega), \quad t>0.
\end{equation}
Here,
\begin{equation}
    \mathcal{I}(x',\omega) = -i\omega \psi(0,x') - \dot{\psi}
    (0,x').
\end{equation}
As usual, the Green's function is composed of solutions to the
frequency domain homogeneous equation (\ref{schrodinger eqn}):
\begin{equation}
    G_\omega(x,x') = \frac{\PsiL (x_<)
    \PsiR
    (x_>)}{W[\PsiL,\PsiR]},
\end{equation}
where $\PsiL$ is the ODE solution that satisfies the outgoing
boundary condition at $x = -\infty$, $\PsiR$ satisfies the
boundary condition at the brane, $x_< = \text{min}(x,x')$, $x_> =
\text{max}(x,x')$, and $W = \PsiL {\PsiR}' - {\PsiL}' \PsiR$ is
the Wronskian. In this section, it is useful to choose the
$x$-coordinate such that $\xb = 0$.  Using the definitions and
results of Sec.~\ref{sec:trapping}, we have
\begin{subequations}
\begin{eqnarray}
    \PsiL(x) & = &
    \begin{cases}
        e^{i\omega x}, & x \rightarrow -\infty, \\
        Re^{i\theta}, & x = \xb \equiv 0,
    \end{cases} \\
    \PsiR(x) & = &
    \begin{cases}
        \cos (\omega x + \delta), & x \rightarrow -\infty, \\
        \eta(\omega), & x = \xb \equiv 0.
    \end{cases}
\end{eqnarray}
\end{subequations}
Because there is no linear derivative in the Schr\"odinger
equation (\ref{schrodinger eqn}), the Wronskian is independent of
$x$ and we can evaluate it at $x = -\infty$
\begin{equation}
    W[\PsiL,\PsiR] =
    -i\omega e^{-i\delta}.
\end{equation}
If we now select the field point to lie on the brane $r = \rb$,
our solution for $\psi$ reads
\begin{equation}\label{psi soln}
    \psib(t) = \int d\omega \, e^{i\omega t} e^{i\delta}
    \eta(\omega) \hat{\mathcal{I}} (\omega), \quad t > 0,
\end{equation}
where
\begin{equation}\label{Ihat}
    \hat{\mathcal{I}}(\omega) = \frac{1}{2\pi} \int
    dx \,
    \PsiL(x) \left[ \psi(0,x) + \frac{1}{i\omega}
    \dot\psi(0,x) \right].
\end{equation}
Note the prominent role played by both the scattering matrix
$S^{1/2} = e^{i\delta}$ and the trapping coefficient $\eta$ in
(\ref{psi soln}).  There are two ways one can evaluate the
$\omega$ integral: either by direct or contour integration.  In
the former, $\eta$ appears directly in the Fourier transform of
the brane signal.  Since $\eta$ is sharply peaked near QBS
resonances, the frequency profile of the brane signal will be
similarly peaked.  On the other hand, if one does the integral by
completing the contour in the upper-half plane, the poles in the
scattering matrix will give the dominant contributions to the
signal; and of those contributions, the QBS poles will be the
longest-lived and hence most important.  Hence, either method
yields the same result.

There are two special cases worth highlighting.  The first
concerns the `scattering experiments' where the initial data is a
coherent compact pulse incident on the brane from the asymptotic
region. We have $\psi(t,x) = f(t-x)$ before scattering, which
allows us to swap $\di_t$ for $-\di_x$ in (\ref{Ihat}). We can
also approximate $\PsiL(x) \approx e^{i\omega x}$. Integrating by
parts gives
\begin{equation}
    \hat{\mathcal{I}}(\omega) = \frac{1}{\pi}
    \int dx \, e^{i\omega x} \psi(0,x).
\end{equation}
Hence, $\hat{\mathcal{I}}$ is the spatial Fourier transform of the
initial data and $|\hat{\mathcal{I}}|^2$ the initial power
spectrum. We can use this to construct the Fourier transform of
the brane signal, keeping in mind that $\psib(t<0)=0$ for this
initial data:
\begin{eqnarray}
    \nonumber \Zb (\omega') & = & \int\limits_{-\infty}^\infty dt \, e^{-i\omega' t}
    \psib(t) = \int\limits_0^\infty dt \, e^{-i\omega' t}
    \psib(t) \\ \nonumber & = & \int\limits_{-\infty}^{\infty} d\omega \,
    \left[ \pi \, \delta(\omega-\omega') +
    \frac{i}{\omega - \omega'} \right] e^{i\delta}
    \eta(\omega) \hat{\mathcal{I}}(\omega).
\end{eqnarray}
To go from the first to second lines, we have used the Fourier
transform of the Heaviside function.  The integral associated with
the first term in square brackets is trivial.  The second integral
expands as:
\begin{equation}\nonumber
    \frac{i}{\pi} \int\limits_{-\infty}^0 dx \,\psi(0,x)
    \int\limits^\infty_{-\infty}  d\omega \frac{e^{i\omega x}
    e^{i\delta} \eta(\omega)}{\omega-\omega'}.
\end{equation}
The pole at $\omega = \omega'$ means that we must evaluate this as
a principal value integral.  Since $x < 0$, we complete the
contour in the lower half plane where the scattering matrix is
analytic ($\eta$ is analytic everywhere).  Assuming the integral
over the semi-circle at infinity vanishes, we obtain
\begin{equation}\label{power spectrum: scattering}
    \Zb (\omega) = 2\pi e^{i\delta} \eta(\omega)
    \hat{\mathcal{I}}(\omega), \quad \frac{dE}{d\omega} = 4\pi^2
    \xi(\omega) |\hat{\mathcal{I}}(\omega)|^2,
\end{equation}
where we have defined the power spectrum in the usual way
$dE/d\omega = |\Zb(\omega)|^2$.  Physically, this says that the
gravity wave power seen on the brane is given by the power in the
initial pulse times $\xi(\omega)$.  In this sense, one could call
$\xi(\omega)$ a `filter-factor', in that it tells us how the
potential processes an input from infinity into a brane signal. Of
course, influence of the QBSs is dominant in this process, because
they give rise to the dominant features in $\xi(\omega)$.

The second special case concerns brane localized initial data:
\begin{eqnarray}\nonumber
    \psi(0,x) & = & A \, \delta(x), \quad \dot\psi(0,x) = B \,
    \delta(x), \\
    \hat{\mathcal{I}}(\omega) & =  &\frac{R e^{i\theta}}{2\pi}
    \left(A + \frac{B}{i\omega}\right).
\end{eqnarray}
A similar argument to before gives the power spectrum
\begin{equation}
    \frac{dE}{d\omega} = \xi(\omega) R^2(\omega) \left[ A^2 +
    \frac{B^2}{\omega^2} \right].
\end{equation}
This is a little more complicated than the scattering experiment
case (\ref{power spectrum: scattering}) in that knowledge of the
trapping coefficient and initial data is not enough to determine
$dE/d\omega$, one also needs to know $R(\omega)$.  But we do know
the behaviour near QBS resonances
\begin{equation}
    \frac{dE}{d\omega} \approx \left[ A^2 +
    \frac{B^2}{\omega^2} \right] \frac{R^4(\omega) \Gamma_n^2}{4
    (\omega - \varpi_n)^2 + \Gamma_n^2}, \quad \omega \approx \varpi_n.
\end{equation}
Hence, the system's resonances manifest themselves as Lorentzian
peaks in the power spectrum.

To summarize: In this section we have constructed the formal
Green's function solution for the brane signal.  Both the
scattering matrix and trapping coefficient appear explicitly in
this solution, implying that the signal will be dominated by QBS
resonances.  The brane power spectra arising from the scattering
of compact pulses from infinity is completely described by the
pulses' Fourier transform and $\xi$.  However, more information is
required for different types of initial data, including the
brane-localized case.  Regardless of the details, the prominence
of $\eta$ within the Green's function ensures that $dE/d\omega$ is
peaked about the ES resonant frequencies for any type of initial
data.

\subsection{Coherent states}

Knowledge of the power in a given signal is not enough to
reconstruct the signal in the time domain; much information is
contained within the phase.  So despite our understanding of the
Green's function of the problem, it is still useful to conduct
numeric scattering experiments to see what kind of behaviour is
possible.

In quantum potential problems featuring discrete wavefunctions,
one can often construct `coherent state' solutions that are wave
packets whose dynamics mimic that of a classical particle
travelling in the same potential.  In the current problem, we know
that far brane configurations have a number of lightly damped QBS
modes.  For example, the $(\gamma,\rb,L) = (20,250,1)$ case has
around seven modes with lifetimes $t_{1/2} \gtrsim 1000$.  On
shorter timescales, these act like discrete bound states.  Hence
we can construct wave packets that behave like classical
particles; for example, they can bounce back and forth within the
resonant cavity.  Of course, as time passes the QBS modes decay
resulting in the decoherence of the pulse.

To see these effects explicitly, we consider a scattering process
in the $(\gamma,\rb,L) = (20,250,1)$ case.  The results of our
numeric integration for $\psi$ are shown in Fig.~\ref{fig:coherent
profile}. Cosine modulated Gaussian initial data is fired at the
potential barrier from well inside the photon sphere: a good
percentage of this is reflected back to the black hole, while the
transmitted part has had a redistribution of the frequencies which
make it up, with preferential selection going to the frequencies
corresponding to maxima of the trapping coefficient $\eta$. After
scattering off the brane, the lowest frequencies in the signal
become effectively trapped between the brane and the barrier, and
bounce between them. While some leakage occurs at each bounce off
the barrier, they cannot penetrate the barrier far, which means
that they persist for a significant number of bounces.  If we were
to increase $L$ in this experiment, we would see the bouncing last
more or less indefinitely since the leakage at each bounce is so
tiny.
\begin{figure}
\includegraphics[width=\columnwidth]{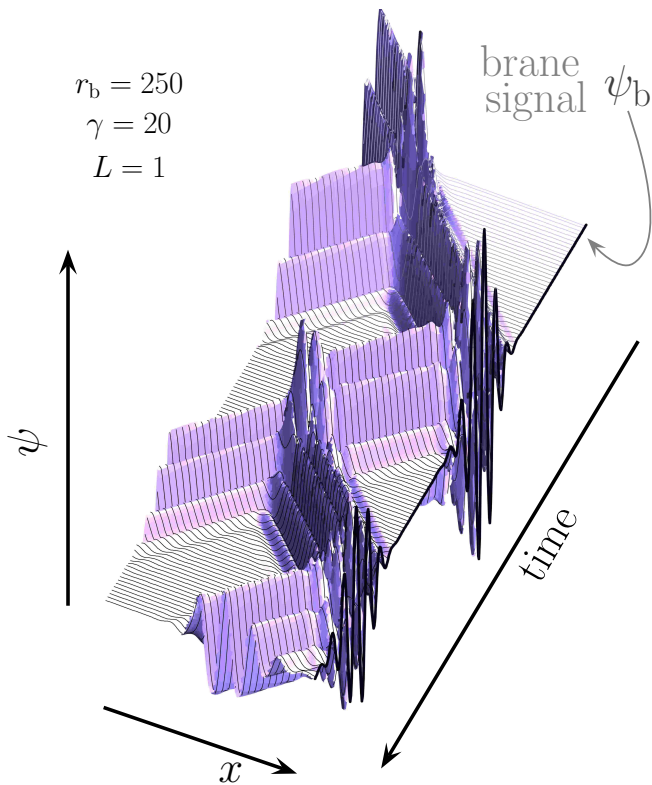}
\caption{A cosine-modulated Gaussian pulse becomes temporarily
trapped in the potential well for the case
$(\gamma,\rb,L)=(20,250,1)$. Initial data is chosen as
$\psi=\cos(x)\exp[-(x-x_0)^2/2\sigma^2]$ centred at $x_0=-20$ with
a variance of $\sigma=5$, moving to the right. In Fourier space
this signal is a Gaussian centred about $\omega=1$. We show the
signal undergoing two reflections off the brane; the potential
barrier at the photon-sphere partially reflects the signal
resulting in a bouncing, temporary trapping of the wave.
\label{fig:coherent profile}}
\end{figure}

\begin{figure}
\includegraphics[width=\columnwidth]{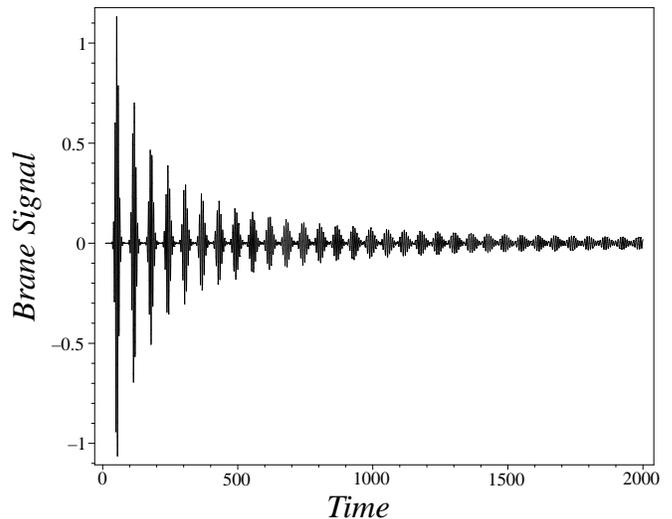}
\caption{The brane signal for the situation shown in
Fig.~\ref{fig:coherent profile}. The gravitons appear in short
bursts in this model, despite arising from a single coherent state
in the bulk. Only at late times does this morph into what looks
like `normal' behaviour for gravity waves on a brane.
\label{fig:coherent signal}}
\end{figure}

To a brane observer things can look pretty strange. If the initial
pulse is particularly narrow, the bouncing of the signal between
brane and barrier would result in a brane signal which disappears
and then reappears many times over---a state which looks like a
collection of recurrent gravitons. This effect is shown in
Fig.~\ref{fig:coherent signal} for a long time integration.

\begin{figure}
\includegraphics[width=\columnwidth]{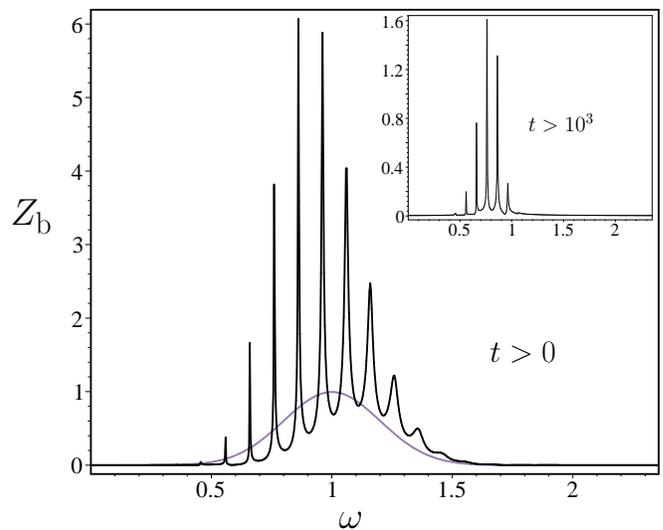}
\caption{Fourier transform of the brane signal in
Fig.~\ref{fig:coherent signal}. The original signal is a Gaussian
in Fourier space (grey curve), which gets combed into discrete
frequencies over successive oscillations between the photon sphere
and the brane. The very late time part of the signal is shown in
the inset: the higher frequency components have a quicker damping
time so only the lowest QNMs are left. \label{fig:coherent FFT}}
\end{figure}

The Fourier transform of this signal, Fig.~\ref{fig:coherent FFT},
shows which modes dominate the waveform. The main curve shown the
FT of the brane signal; the Gaussian is the FT of the initial
data, and the inset shows the signal at very late times. We can
easily discern the peaks in the spectrum which correspond to the
spikes in $\eta$, shown in the inset of Fig.~\ref{fig:varying L}.
Indeed, the signal in Fourier space looks very like
$\text{(initial data)} \times \eta$, as expected from our Green's
function analysis. Note, however, the the lowest modes are far too
narrow to resolve numerically ($\sim10^{-10}$).

\subsection{Steady state gravitational radiation: brane-black hole thermal
equilibrium}

Our last `initial value problem' example does not really involve
initial data at all.  Consider a situation where the gravity waves
near the horizon are a super position of standing waves:
\begin{equation}
    \psi(t,x) = \int d\omega\,Z_\infty(\omega) e^{i\omega t} \cos(\omega x +
    \delta), \quad x \rightarrow \infty.
\end{equation}
This represents a situation where there is an equal amount of
energy flux from and to the black hole.  Physically, this is a
good model of a brane in thermal equilibrium with the flux of
Hawking radiation coming from the horizon.  In that case, the
power in the asymptotic waveform is given by the blackbody formula
\begin{equation}
    |Z_\infty(\omega)|^2 = \frac{\omega^3}{e^{\omega/T_\text{H}} -
    1},
\end{equation}
where the Hawking temperature is $T_\text{H} = \kappa/2\pi$ and
$\kappa$ is the surface gravity.  Somewhat trivially, we use the
definition of the trapping coefficient to obtain the brane signal
\begin{equation}
    \psib(t) = \int d\omega \, \eta(\omega) Z_\infty(\omega)
    e^{i\omega t},
\end{equation}
and power spectrum
\begin{equation}
    \frac{dE}{d\omega} = 4\pi^2 \xi(\omega) |Z_\infty(\omega)|^2 =
    \frac{4\pi^2 \xi(\omega) \omega^3}{e^{\omega/T_\text{H}} - 1}.
\end{equation}
This result is pretty much identical to what we had from the
Green's function answer for the scattering of a compact pulse: the
power of the brane signal is the power in the waveform at infinity
times $\xi(\omega)$.  Hence, the same conclusion holds here: The
blackbody spectrum is amplified at the positions of QBS
frequencies, and that amplification has a Lorentzian lineshape
whose width is proportional to the lifetime of the resonance.

\section{Discussion}\label{sec:conc}

\subsection{Summary of resonances}

\begin{figure}
\includegraphics[width=\columnwidth]{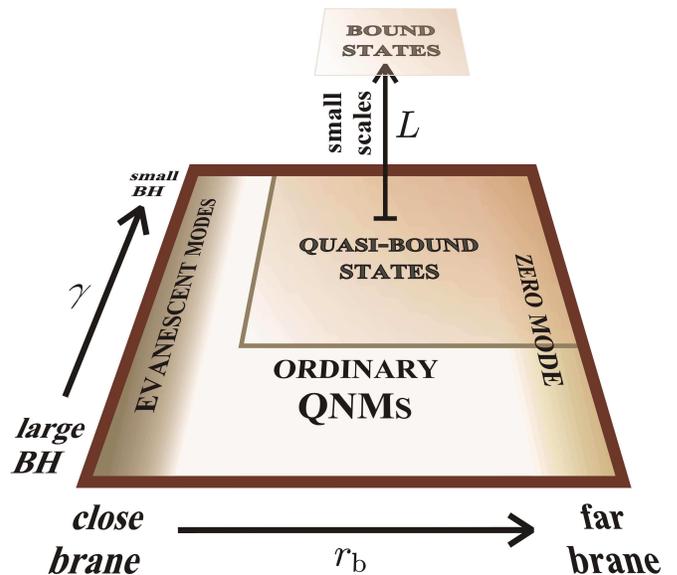}
\caption{Dominant resonant behaviour in different regions of the
$(\gamma,\rb,L)$ parameter space} \label{fig:parameter 3D}
\end{figure}

One of the main objectives of this paper was to gain complete
knowledge of the quasi-normal resonances of Einstein-static
braneworlds.  Practically, this meant calculating QNM frequencies
in different regions of the 3-dimensional $(\gamma,\rb,L)$
parameter space characterizing the bulk gravity waves.  Using two
different numerical methods and a variety of analytic techniques,
we managed to determine the behaviour of the resonant modes in a
number of different interesting limits.  In
Fig.~\ref{fig:parameter 3D}, we have summarized the key
qualitative features of the resonances as a function of the model
parameters.

Another way of presenting our results involves classifying
resonances as over-damped ($\Imag\omega \gg \Real \omega$),
moderately damped ($\Imag\omega \sim \Real\omega$), or
under-damped ($\Imag\omega \ll \Real\omega$). Then we can ask:
Which parameter choices produce which type of resonance?
\begin{description}

\item{\bf Over-damped modes:} Evanescent resonances occur when the
brane is very close to the horizon -- gravitons are sucked off the
brane with great efficiency. In fact, when the brane is extremely
close the concept of gravity waves is a bit dubious: all QNMs are
purely imaginary, so no oscillations take place at all, and such
modes just decay exponentially. For branes a bit further out there
may be a small number QNMs with non-zero real part.

\item{\bf Moderately damped modes:} When the brane is located at a
moderate distance from the black hole, or the black hole is large
so that no resonant cavity exists, we have the situation of
`normal' QNMs. This means that any signal will typically be
dominated by the mode with the lightest damping~-- which also has
the lowest frequency. This mode tends to dominate any signal very
quickly.  As discussed in detail in our previous work
\cite{Seahra:2005us}, all the resonant modes of pure tension
branes are in this class.

\item{\bf Under-damped modes:}  Very long-lived modes arise in
several different situations:

\begin{description}

\item{\it Zero-mode:} For all $(\gamma,L)$, as
$\rb\rightarrow\infty$ the fundamental mode becomes purely real,
and matches the GR frequency; this mode is stuck to the brane. The
higher overtones remain complex.

\item{\it Quasi-bound states:} When the black hole is small
($\gamma \gtrsim 2$) there exists a well in the potential;
provided the brane is located past this well a pseudo-resonant
cavity forms and QNMs with $\varpi^2$ less than the peak in the
potential may be interpreted as quasi-bound states. These modes
have tiny imaginary parts implying a very long lifetime before
they tunnel into the horizon.

\item{\it Bound states:} Quasi-bound states become bound on small
scales: as $L\rightarrow\infty$, we have $\text{Im}(\omega)$
vanished exponentially quickly. This implies that on small scales
an ES brane will look like a two brane model with virtually even
spacing between all modes except the fundamental.

\end{description}

\end{description}

\noindent This list and Fig.~\ref{fig:parameter 3D} encapsulate
our essential results on the quasi-normal modes of Einstein-static
braneworlds.

\subsection{The key role of the trapping coefficient in the initial value problem}

An intriguing, and somewhat unexpected, by-product of our work has
been the trapping coefficient $\eta$.  Recall that this was
originally introduced to help us calculate quasi-bound state
frequencies when $\rb$ became too large for the series method to
handle.  But in our Green's function analysis of the initial value
problem in Sec.~\ref{sec:greens}, $\eta(\omega)$ played a
prominent role Fourier transform of the brane signal.  Indeed, for
the case of a compact pulse propagating towards the brane from
infinity, the power spectrum on the brane was shown to be simply
$\xi = \eta^2$ times the power spectrum of the pulse.  In other
words, the trapping coefficient tells us to what degree each
Fourier component from $x = -\infty$ is localized on the brane.

In this sense, the precise functional form of $\xi(\omega)$ is
crucial.  To see how, assume for a moment that we only know the
quasi-normal modes of the system and are ignorant of
$\xi(\omega)$. Then, in any given scattering experiment we would
predict that the brane power spectrum will be enhanced in some
range $\Delta \omega \sim \Gamma_n$ around each resonant frequency
$\varpi_n$; but we would not know beforehand to what degree each
mode would be excited by the initial data.  This is a known
conundrum in black hole perturbation theory
\cite{Andersson:1996cm}: quasi-normal modes by themselves only
tell half the story, one also needs their `excitation strengths'
to really make detailed predictions.

However, using $\xi$ to predict brane signal side-steps the whole
issue because information about the excitation strengths is
\emph{already} encoded in the heights of the Lorentzian peaks
representing the resonances. In small black hole/far brane cases
$\xi$ is a very spiky function, which means that it effectively
combs initial data into a sum of lightly damped sinusoids centered
about quasi-bound state frequencies, and efficiently suppresses
the power in-between. Even when the QNMs are heavily damped the
trapping coefficient plays an important role, amplifying signals
in the region near the fundamental QNM, but suppressing signals of
lower frequency.  At the other end of the spectrum, $\xi$ tells us
in detail how high energy waves from the horizon are felt on the
brane, a subject on which quasi-normal modes have nothing to say.

But there are limitations to what can be learned in this fashion.
If initial data has support outside the asymptotic region (i.e.,
close to the brane), things become muddled because our simple
Green's function result for the scattering of distant pulses does
not apply. The only sure thing in these cases comes from the
quasi-normal modes, in that they still predict the position and
width of resonant peaks in the brane signal.  The actual power
carried in each mode must be obtained by other means. Having
stated that caveat, the trapping coefficient remains a powerful
tool for describing the resonances of the Einstein-static
braneworld. It has the attractive quality of being much easier to
calculate than QNMs from a series solution, and it neatly
communicates most of the important features of the scattering
problem.  We expect it will be of considerable use in other
`one-sided' brane scattering problems that feature lightly damped
resonances and a flat asymptotic region.

\subsection{Implications for cosmology and future work}

Einstein-static branes become closed cosmological branes if we
allow their radii to depend on time $\rb = a(t)$.  The dynamics of
$a(t)$ is governed by the Friedman equation (\ref{friedman}).  The
evolution of gravity waves in the most general scenarios is more
complicated that the static case considered in this paper, with a
formal solution of the form
\begin{equation}
    \psib(t) = \int dt' \, dx'\, G(t,\xb;t',x') \mathcal{I}(t',x').
\end{equation}
Here, $G$ is the real space retarded Green's function and
$\mathcal{I}$ represents a generic source which includes
$\psi(t,x)$ and perhaps other fields. Solving integro-differential
equations of this type are the essential obstacle in braneworld
cosmology \cite{Maartens:2003tw,Mukohyama:2001yp}.  This is
largely because the moving brane boundary condition makes it
exceedingly difficult to deal with the Green's function in generic
situations.

However, in the high-frequency regime we can make use of a
multiple-scales approximation \cite{Battye:2004qw} to make some
progress.  This relies on the fact that if gravitational waves
oscillate much faster than the brane moves, they will `see' the
brane to be stationary.  To zeroth order in this approximation,
the Green's function above is just the frequency space Green's
function we developed in Sec.~\ref{sec:IVP} transformed into the
time domain. On dimensional grounds, `slow' brane motion implies
$\omega \gg H$, which means that our static results should apply
to waves with oscillation periods much shorter than the Hubble
time. Another way to state this is that the static Green's
function is a good approximation to the moving Green's function
provided $H(t-t') \ll 1$.

Hence, any quasi-normal modes we have found will be approximate
resonant solutions to the moving brane problem if $\Real\omega_n
\gg H$. In this picture, one thinks of the frequency of such modes
as quantities evolving in time in a quasi-static, adiabatic
process. Then, plots like Figs.~\ref{fig:ribcage} and \ref{fig:far
brane qnm}, which show the variation of $\omega_n$ with $\rb$, can
be re-interpreted as depicting the evolution of resonant
frequencies with cosmological epoch, provided we only pay
attention to the high frequency modes. Ideally, we would like to
understand the systems resonances at lower frequencies, which
would involve moving beyond the zeroth order static approximation
to the the Green's function.  An interesting future project would
involve formalizing such an approximation scheme and then defining
generalized, \emph{dynamic} quasi-normal modes.

Finally, we note that due to the lack of time symmetry, it is more
common in cosmology to characterize modes by their comoving
wavelengths as opposed to their frequencies.  For branes in the
pure tension or far brane regimes, all resonances we found had
$\Real\omega_n > k$, as might be expected from the null momentum
condition in 5 dimensions.  Hence, our slow motion approximation
can be re-cast in terms of spatial wavelengths: \emph{on
sufficiently sub-Hubble scales, gravity waves on a cosmological
brane will be dominated by the quasi-normal resonances of the
corresponding static brane}. Most importantly, this means that we
recover a RS-like zero-mode for late cosmological epochs and on
small scales.  This is a critical test for the viability of
braneworld cosmological modes involving a bulk black hole -- if it
were not true, we would see significant departures from general
relativity over astrophysical distances in the current universe.
However, it is tantalizing that we only find an \emph{approximate}
zero mode. This leaves the door open to finding small corrections
to standard gravity induced by the presence of a bulk black hole.
This is an important subject for future study.

\begin{acknowledgments}

CC is supported by PPARC and SSS is supported by NSERC.  We would
like to thank K Koyama, A Mennim, D Wands and especially R
Maartens for many productive discussions on this work.

\end{acknowledgments}

\appendix

\section{The close brane limit and evanescent
modes}\label{sec:close brane}

In this appendix, we find an analytic solution to the master wave
equation when the brane is close to the black hole horizon $\rb
\sim 1$.  The goal is to obtain an exact expression for the purely
imaginary evanescent modes seen in Sec.~\ref{sec:QNM results} for
$\rb \rightarrow 1$.

The extreme close brane limit is defined by $\rb - 1 \ll 1$. Since
$r \in (1,\rb)$, we have
\begin{equation}
    f \approx 2 \kappa (r-1) \approx e^{2\kappa x} \ll 1,
\end{equation}
in between the brane and the black hole horizon (recall that
$\kappa = (2+\gamma^2)/\gamma^2$ is the surface gravity). We
expand the wave equation (\ref{master wave pde}) to leading order
in $e^{2\kappa x}$, and apply a linear coordinate transformation
$X = 2\kappa(\xb-x)$ to simplify matters. Assuming $\psi_k =
e^{i\omega t}\Phi_{k\omega}(X)$, and defining the parameters
\begin{eqnarray} \nonumber
    \Omega & = & \frac{\omega}{2\kappa}, \quad \Delta =
    \frac{3}{2} \left( 1 - \frac{1}{\rb} \right), \\ \beta^2 & = &
    \frac{[6 + (L^2 + 2L + 3)\gamma^2](\rb-1)}{2(2+\gamma^2)},
\end{eqnarray}
yields the simple wave equation
\begin{subequations}
\begin{eqnarray}\label{close brane ODE}
    \Omega^2 \Phi_{k\omega}(\x) & = & -\Phi_{k\omega}''(\x) + \beta^2 e^{-\x} \Phi_{k\omega}(\x),
    \\ \label{close brane BC} \Phi_{k\omega}'(0) & = & -\Delta \Phi_{k\omega}(0).
\end{eqnarray}
\end{subequations}
Note that the $\x$ coordinate has been selected such that the
brane is at $\x = 0$ and the horizon is at $\x = \infty$.  The ODE
(\ref{close brane ODE}) has the following exact solution
satisfying (\ref{close brane BC}):
\begin{equation}
    \Phi_{k\omega}(\x) = F_1 I_\nu(2\beta e^{-\x/2}) +
    F_2 K_{-\nu}(2\beta e^{-\x/2}),
\end{equation}
where $\nu = 2i\Omega$ and
\begin{subequations}
\begin{eqnarray}
    F_1 & = & \beta K_{1-\nu}(2\beta)+(i\Omega + \Delta)
    K_\nu(2\beta),
    \\ F_2 & = & \beta I_{1+\nu}(2\beta)+(i\Omega - \Delta)
    I_\nu(2\beta).
\end{eqnarray}
\end{subequations}
Here, $I_\mu$ and $K_\mu$ are the modified Bessel functions of
order $\mu$.  In order to find the quasi-normal frequencies, we
need to know about the asymptotic behaviour of these Bessel
functions near the horizon at $\x = \infty$.  From the relations
\cite{Arfken}
\begin{subequations}
\begin{eqnarray}
    \label{I expansion} I_\mu(2z) & = & \frac{z^\mu}{\mu!}
    \left[ 1 + \frac{z^2}{\mu+1} + \mathcal{O}(z^4) \right], \\
    K_\mu(2z) & =  &\frac{\pi}{2} \frac{I_{-\mu}(2z)-I_\mu(2z)}{\sin
    \pi\mu},
\end{eqnarray}
\end{subequations}
we deduce the large $\x$ behaviour
\begin{subequations}
\begin{eqnarray}
    I_\nu(2\beta e^{-\x/2}) & \rightarrow & \text{const.}\times
    e^{-i\Omega\x}, \\ K_{-\nu}(2\beta e^{-\x/2}) & \rightarrow & \text{const.}\times
    \sin(\Omega\x+\varphi),
\end{eqnarray}
\end{subequations}
where $\varphi$ is some constant (real) phase.  We see that the
former represents a purely outgoing wave at the horizon, while the
latter is a mixture of incoming and outgoing radiation.  To have a
QNM, we need to set the contribution from the incoming radiation
equal to zero.  Hence, the quasi-normal frequencies are the
(complex) $\omega = 2\kappa\Omega$ for which
\begin{equation}\label{close brane QNM cond'n}
    F_2 = F_2(\omega;\gamma,\rb,L) = 0.
\end{equation}

Notice that $\rb - 1 \ll 1$ means that $\beta$ is a small
parameter.  Furthermore, for any given choice of $\gamma$ and $L$,
we see that $\Delta$ is of the same order as $\beta^2$. Therefore,
it makes sense to expand $F_2$ in powers of $\beta$.  Again using
(\ref{I expansion}), we find
\begin{equation}
    F_2 = \beta^\nu \left[ \frac{i\Omega}{\nu!} +
    \frac{\beta^2(1+i\Omega)}{(1+\nu)!} - \frac{\Delta}{\nu!} +
    \mathcal{O}(\beta^4) \right].
\end{equation}
To leading order in $\beta$, we find that $F_2 = 0$ implies
$\Omega/\nu! = 0$, which has solutions
\begin{equation}
    \Omega = 0, \quad \nu = -1,-2,-3\ldots
\end{equation}
Expressing $\Omega$ and $\nu$ in terms of $\omega$ and $\kappa$
yields:
\begin{equation}\label{close brane QNMs}
    \lim_{\rb \rightarrow 1} \omega_n = i \kappa n, \quad n = 0,1,2,3\ldots.
\end{equation}
This simple formula holds for all $\gamma$ and $L$.  So, when the
brane is very close to the horizon, the quasi-normal frequencies
are evenly spaced along the positive imaginary axis, in agreement
with the results obtained from the series solution in
Sec.~\ref{sec:QNM results}. It is interesting to see that the QNM
frequencies scale with the surface gravity of the black hole in
this limit, just like the small-$\gamma$ QNM frequencies of a
S-AdS black hole \cite{Horowitz:1999jd} or the overtones of a pure
tension Einstein-static brane \cite{Seahra:2005us}.  Finally, we
have quantitatively compared (\ref{close brane QNMs}) to our
series solutions for the cases shown in Fig.~\ref{fig:ribcage} and
found excellent agreement.

\bibliography{ES_paper}

\end{document}